\documentclass[a4paper]{article}
\pdfoutput=1 % if your are submitting a pdflatex (i.e. if you have
\usepackage{jcappub} % for details on the use of the package, please
\usepackage{multirow}

\usepackage[T1]{fontenc} % if needed

\usepackage{dcolumn}% Align table columns on decimal point
    \newcolumntype{d}[1]{D{.}{\cdot}{#1}}
    \newcolumntype{.}{D{.}{.}{-2}}
    \newcolumntype{,}{D{,}{,}{-3}}

\usepackage{booktabs}

\title{Updated constraints on amplitude and tilt of the tensor primordial spectrum}

\author[a,b,1]{Giacomo Galloni,\note{Corresponding author.}}
\author[c,d,e]{Nicola Bartolo,}
\author[c,d,e,f]{Sabino Matarrese}
\author[a,b]{Marina Migliaccio}
\author[c,d]{Angelo Ricciardone}
\author[a,b]{and Nicola Vittorio}

\affiliation[a]{Dipartimento di Fisica, Università di Roma Tor Vergata,\\Via della Ricerca Scientifica, 1, 00133, Roma, Italy}
\affiliation[b]{INFN, Sezione di Roma 2,\\Via della Ricerca Scientifica, 1, 00133 Roma, Italy}
\affiliation[c]{Dipartimento di Fisica e Astronomia ``G. Galilei'', Universit\`a degli Studi di Padova,\\Via Marzolo 8, I-35131 Padova, Italy}
\affiliation[d]{INFN, Sezione di Padova,\\Via Marzolo 8, I-35131 Padova, Italy}
\affiliation[e]{INAF - Osservatorio Astronomico di Padova,\\Vicolo dell’Osservatorio 5, I-35122 Padova, Italy}
\affiliation[f]{Gran Sasso Science Institute,\\Viale F. Crispi 7, I-67100 L’Aquila, Italy}

\emailAdd{giacomo.galloni@roma2.infn.it}
\emailAdd{nicola.bartolo@pd.infn.it}
\emailAdd{sabino.matarrese@pd.infn.it}
\emailAdd{marina.migliaccio@roma2.infn.it}
\emailAdd{angelo.ricciardone@pd.infn.it}
\emailAdd{nicola.vittorio@uniroma2.it}

\abstract{We have taken a comprehensive approach to update the limits on the tensor-to-scalar ratio ($r$) and the tensor spectral index ($n_t$), using 10 datasets from the BICEP/Keck Array 2015 and 2018, \textit{Planck} releases 3 and 4, and LIGO-Virgo-KAGRA Collaboration. By fitting the complete $\Lambda$CDM+$r$+$n_t$ model with two different approaches for the tensor sector, we have not only established which method is the most reliable, but have also achieved the strongest constraint on the tensor-to-scalar ratio in current literature: $r<0.028$ and $-1.37 < n_t < 0.42$ at 95\% confidence level. Furthermore, our examination of the common signal detected by the NANOGrav Collaboration further confirms that a simple power-law cannot reconcile the constraints from different datasets if the NANOGrav detection is due to a primordial inflationary gravitational wave background, as previously shown in the literature.}
\arxivnumber{2208.00188}

\begin{document}

\maketitle

\flushbottom

\section{Introduction}\label{sec: intro}

Tensor perturbations in Cosmology can be produced in the very early Universe along with scalar and vector ones, and constitute a background of Gravitational Waves (GWs)\cite{Guzzetti:2016mkm}. These are known to source the divergenceless component of the Cosmic Microwave Background (CMB) polarization, the B-mode, with a maximum contribution on large scales, where they are boosted by rescattering onto electrons freed by cosmic reionization, and, on the degree-angular scale, corresponding to the cosmological horizon at recombination \cite{Seljak_1997, Kamionkowski_1997, Kamionkowski_2016}.

Due to the insight into the physics of the very early Universe that the discovery of a stochastic GW background would give, this polarization is the main target of ongoing and upcoming experiments like BICEP/Keck Array (BK) \cite{Ade_2014}, Simons array \cite{Suzuki_2016}, Simons Observatory \cite{Ade_2019}, Stage-IV \cite{abazajian2016cmbs4} and the Light satellite for the study of B-mode polarization and Inflation from cosmic microwave background Radiation Detection (LiteBIRD) \cite{Hazumi:2019lys, PTEP_LiteBIRD}.

The goal of a full analysis of CMB photon anisotropies would not only be to assess the standard $\Lambda$ $+$ Cold Dark Matter ($\Lambda$CDM) model and its six parameters \cite{Planck_parameters, Planck_like}, but also two additional ones which characterize the B-mode polarization \cite{Planck_2018} (\textit{Planck} 2018). These two parameters, typically are a tensor-to-scalar ratio $r$, which parametrizes the amplitude of primordial tensor perturbations at some reference pivot scale, and a spectral tilt $n_t$, parametrizing the scale dependence of such a primordial spectrum. In other words, the tensor spectrum is usually parametrized as a power-law, equivalently to what is done for scalars:
\begin{equation}\begin{aligned}
    P_t(k) &= r A_s \qty(\frac{k}{k_*})^{n_t}\ ,\\
    P_s(k) &= A_s \qty(\frac{k}{k_*})^{n_s-1}\ , 
\label{eq: power_spectra}
\end{aligned}\end{equation}
where $r = A_t/A_s$ is the tensor-to-scalar perturbation ratio, $A_s$ ($A_t$) is the amplitude of scalar (tensor) perturbations, $k_*$ is some pivot scale and $n_s$ ($n_t$) is the scalar (tensor) spectral index.

\begin{figure}[t]
    \centering
    \includegraphics[width=.5\hsize]{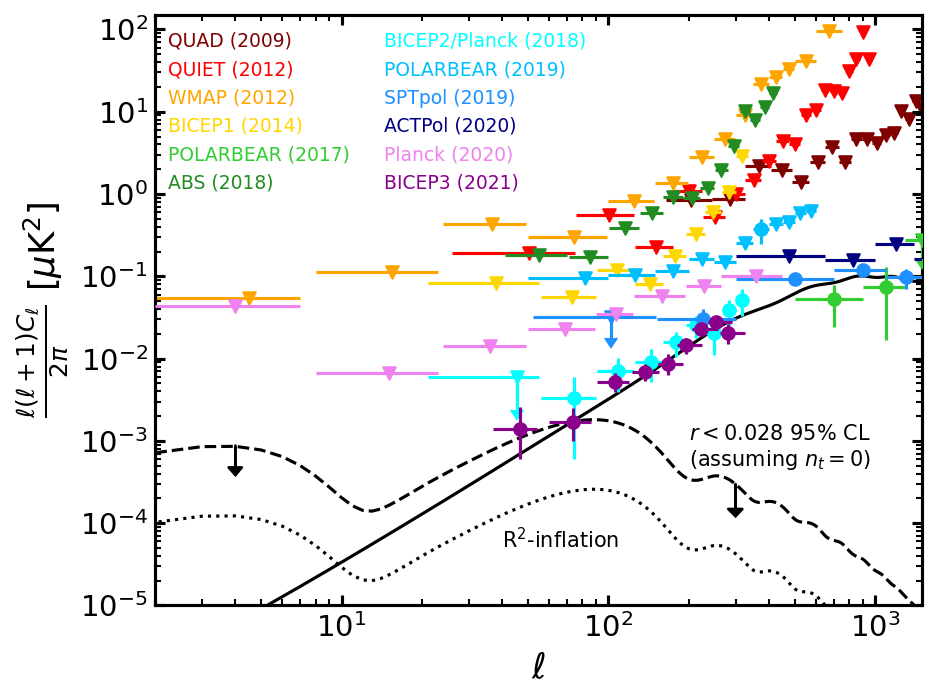}
    \caption{CMB B-mode measurements: BICEP3/Keck Array \cite{BICEP_2021}, \textit{Planck}'s satellite \cite{Tristram2021},ACTPol \cite{Choi_2020}, SPTpol \cite{Sayre_2020}, POLARBEAR \cite{POLARBEAR_2017, POLARBEAR_2019}, BICEP2/Keck Array \cite{Ade_2018}, ABS \cite{Kusaka_2018}, BICEP1 \cite{bicep_1}, WMAP \cite{wmap_2013}, QUIET \cite{QUIET_2012} and QUaD \cite{Brown_2009}. The solid line represents the lensing signal, whereas the dashed and dotted ones are respectively the primordial signal obtained assuming scale invariance and $r = 0.028, 0.004$, the former being the 95\% CL upper bound of this work, assuming scale-invariance, and the latter the prediction of the Starobinski model \cite{Starobinsky}.}
    \label{fig: exp}
\end{figure}

Currently, we have quite a zoology of measurements of CMB B-mode, some of which are shown in figure \ref{fig: exp}, however, CMB experiments as \textit{Planck} are mostly sensitive to the large-scale part of the primordial B-mode spectrum (at least assuming a tensor spectral tilt close to scale invariance). The small scales are hidden below several orders of magnitude of the lensing B-mode as a result of photons passing through the gravitational potential of large-scale structure (compare the dashed, or dotted, line of figure \ref{fig: exp} with the solid one).
Thus, these measurements cannot constrain small scales very well and tend to favor higher values of $n_t$ (blue tilts), providing loose upper bounds on this parameter. This is why CMB experiments are typically flanked to some other small-scale measurement when they are asked to give bounds on the spectral tilt. The quintessential example is GW interferometers, such as LIGO-Virgo-KAGRA (LVK) \cite{aLIGO, aVirgo, ligo-virgo} since they can directly probe the same GWs one tries to study through the B-mode, but at completely different scales.

Although \citet{Tristram:2022} reports the tightest bound on the tensor-to-scalar ratio ($r_{0.05}<0.032$ at 95\% CL), here we want to consider the case in which the tensor spectral tilt is left to vary, beyond the usual consistency relation for single-field slow-roll models ($r=-8 n_t$).
So, while keeping in mind the work by \citet{Tristram:2022}, the actual state-of-the-art on $\qty(r, n_t)$ are the bounds set by \textit{Planck} 2018: $r_{0.01} < 0.066$ and $-0.76 < n_t < 0.52$ at $95\%$ CL, when including both CMB and GWs interferometers (see section \ref{sec: state} for further details on how these bounds are obtained). Here the subscript ${0.01}$ indicates the pivot scale that is typically assumed in this context, i.e. $0.01$ Mpc$^{-1}$. The main goal of this paper is to update these constraints, exploiting newly available data, both from an electromagnetic and a GW perspective. 

Indeed, we study how adding the new data released by BK \cite{BICEP_2021}, \textit{Planck} \cite{Tristram2021} and the LVK collaboration \cite{Abbott_2019_ligo, ligo_meas} can improve our knowledge about the tensor sector. 
\begin{figure}[t]
    \centering
    \includegraphics[width=\textwidth]{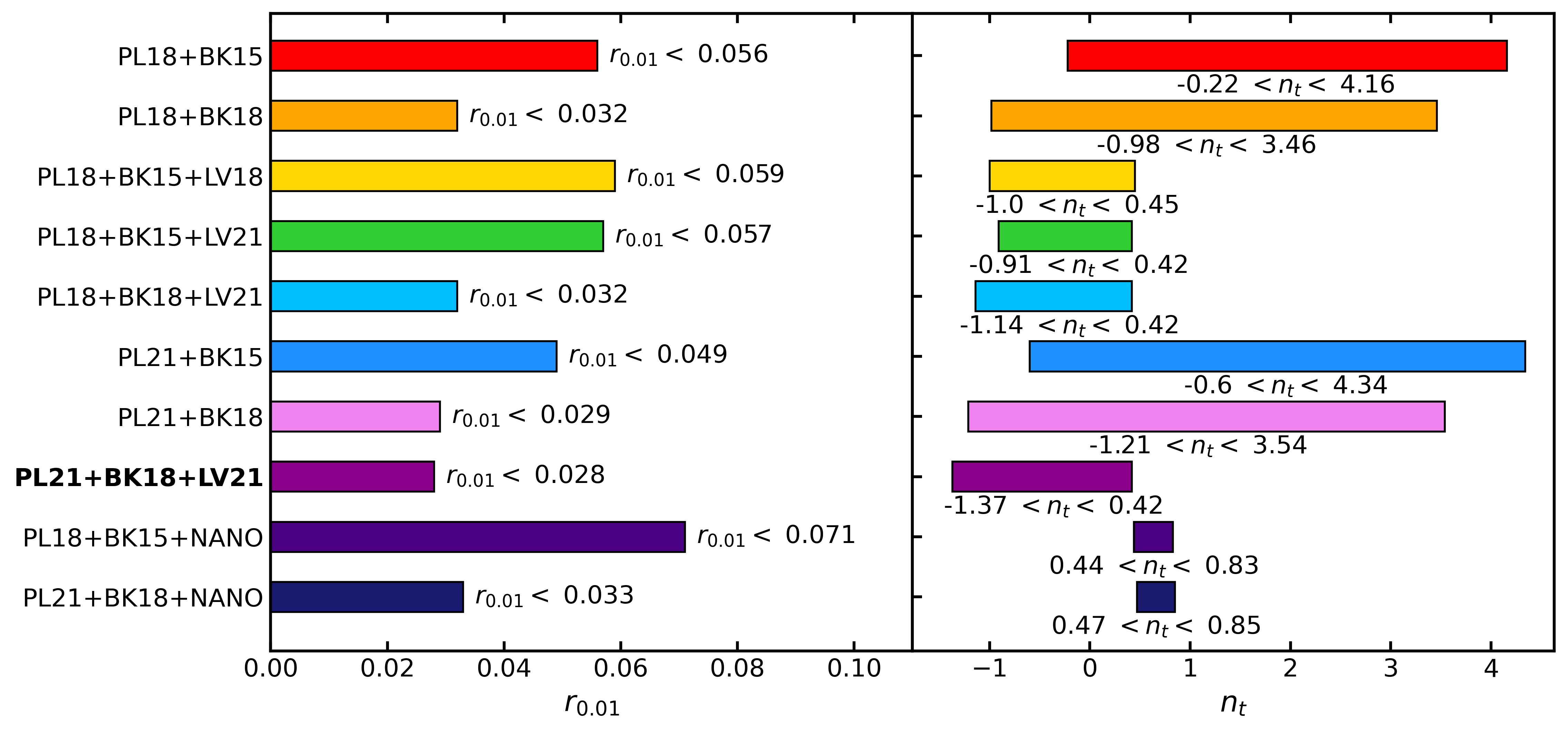}
    \caption{$95\%$ CL intervals for $r_{0.01}$ and $n_t$, considering different datasets, given in table \ref{tab: names}. Our main result is PL21$+$BK18$+$LV21.}
    \label{fig: bounds}
\end{figure}

We perform this analysis exploiting two different approaches on how to sample the tensor sector: the first one is to explore directly the parameter space $\qty(r_{0.01}, n_t)$ with uniform priors \cite{Cabass_2016}, while the second consists of considering $r_{0.01} $ and $ n_t$ as derived parameters from two functions of two tensor-to-scalar ratios at arbitrary scales $k_1$ and $k_2$ \cite{Planck15_inflation, Planck_2018}. We explore the strengths and weaknesses of these two approaches, finding that the former can provide more reliable bounds. The results are summarized in figure \ref{fig: bounds}, where the case PL21$+$BK18$+$LV21 (see section \ref{sec: data}) represents the main result of this work.

Furthermore, we study with the same approach the consequences of newly released data from the North American Nanohertz Observatory for Gravitational Waves (NANOGrav) Collaboration \cite{nanograv_meas}. Indeed, pulsar timing array experiments can be used to detect GWs through the spatial correlation they generate on a network of radio pulses (typically millisecond pulsars). \citet{nanograv_meas} show strong evidence for a common stochastic process affecting several pulsars, however, the spatial quadrupolar correlation (see \cite{Hellings_Downs}), necessary to finally confirm the detection of a stochastic GW background is not convincing. Despite this, if the cosmological nature of this signal is found to be true, this seems to suggest the need to go beyond a simple power-law description of the primordial tensor spectrum, consistently with what has already been found in the literature \cite{Vagnozzi_2020, Kawasaki_2021, Fujita_2022, Benetti_2022}.

This work is organized as follows: in section \ref{sec: MCMC} we give the details on the Markov-Chain Monte-Carlo (MCMC) framework we employ to perform the analysis, while we test the robustness of the two approaches mentioned above for the tensor sector. In section \ref{sec: data} we go through the available data to constrain the tensor spectrum, while in section \ref{sec: state} we give a very brief review of how the state-of-the-art bounds have been obtained. Finally, in section \ref{sec: Results} we report our new bounds for $\qty(r_{0.01}, n_t)$ using 10 different combinations of datasets and we briefly discuss the cases in which we include NANOGrav data. In appendix \ref{sec: LCDM} we provide the posterior distributions of 4 selected combinations of datasets on the six $\Lambda$CDM parameters, comparing the results with the literature. Then, in appendix \ref{sec: coords} we explore an alternative route to test one of the two approaches used in this analysis, while in appendix \ref{sec: TSA} we repeat the analysis leading to section \ref{sec: Results}, but using the approach of \cite{Planck15_inflation, Planck_2018}.

\section{MCMC analysis} \label{sec: MCMC}

Let us report some details on the technique used to extract these constraints from the data, i.e. the MCMC analysis \cite{MCMethod, Ulam, MonteCarlo, Lewis:2002ah, Lewis:2013hha}. We use \texttt{Cobaya} \cite{Torrado:2020xyz} to run our MCMC chains, whose results are \footnote{\url{https://github.com/CobayaSampler/cobaya}.}analyzed through \texttt{GetDist}\footnote{\url{https://github.com/cmbant/getdist}.} \cite{Lewis:2019}, and \texttt{CAMB} \footnote{\url{https://github.com/cmbant/CAMB}.}\cite{Lewis:2013hha, Lewis_2000, Howlett_2012} to generate the CMB spectra.

\subsection{\texorpdfstring{$\Lambda$}{TEXT}CDM parameters}

For what regards the 6 $\Lambda$CDM parameters $\qty{A_s,\ n_s,\ \Omega_bh^2,\ \Omega_{cdm}h^2,\ \theta_s,\ \tau_{reio}}$, which are not the focus of this work, we only mention that we choose uniform and very wide priors, as it is usually done in order to include the fewer information as possible. In fact, it is important to let them vary to capture completely the variance of the tensor parameters \cite{Tristram:2022}. In particular, Tab.\ref{tab: priors} reports the ranges of the uniform priors we choose, together with the usual meaning of the parameters. Note that, if not said otherwise, these parameters are always kept free to vary.
\begin{table}[t]
\centering
\begin{tabular}{cc}
\toprule
\textbf{Parameter}& \textbf{Prior Range}\\ \midrule
$log(10^{10}A_s)$ & ${[}1.61, 3.91{]}$  \\
$n_s$             & ${[}0.8, 1.2{]}$    \\
$\Omega_b h^2$    & ${[}0.005, 0.1{]}$  \\
$\Omega_{cdm} h^2$& ${[}0.001, 0.99{]}$ \\
$\theta_s$        & ${[}0.5, 10{]}$     \\
$\tau_{reio}$     & ${[}0.01, 0.8{]}$   \\ \midrule
$r_{0.01}$        & ${[}10^{-5}, 3{]}$  \\
$n_t$             & ${[}-5, 5{]}$  \\
$r_1$             & ${[}0, 3{]}$ \\
$r_2$             & ${[}0, 3{]}$ \\ \bottomrule
\end{tabular}
\caption{\label{tab: priors}Prior ranges for the $\Lambda$CDM parameters $+$ the tensor sector. Here $A_s, n_s$ are the amplitude and the tilt of the scalar primordial perturbations, $\Omega_b, \Omega_{cdm}$ are the energy densities of baryons and cold dark matter, $h$ is the Hubble constant $H_0$ divided by 100, $\theta_s$ is the angular scale of the sound horizon at recombination and $\tau_{reio}$ is the optical depth of reionization. Instead, $r_{0.01}$ is the tensor-to-scalar ratio at $0.01$ Mpc$^{-1}$, $n_t$ is the tensor spectral tilt and $r_1, r_2$ are two tensor-to-scalar ratios at two arbitrary scales $(k_1, k_2) = (0.002, 0.02)$ Mpc $^{-1}$.}
\end{table}
For further details on how the newly available data affect this sector of parameters, see appendix \ref{sec: LCDM} and \cite{Planck_parameters, Tristram:2022, Couchot_2017a, Couchot_2017b}.

\subsection{Tensor parameters}

For what concerns the tensor sector of the parameters, we must give some more details.
As mentioned in section \ref{sec: intro}, the tensor power spectrum is customarily parametrized as a power law around an arbitrary pivot scale $k_*$
\begin{equation}
    P_t(k) = A_s r \qty(\frac{k}{k_*})^{n_t}\ .
    \label{eq: tensor_spectrum}
\end{equation}
Naively, one would choose a uniform prior for both $r$ and $n_t$, such as $0<r<3$ and $-5<n_t<5$, however, we have to keep in mind that we actually have not detected $r$ yet. Thus, for very low values of the amplitude, $n_t$ will be completely free to vary in its prior ranges, producing a pathological behavior of the final posteriors.

In a Bayesian context, there are two ways to solve the issue (for an alternative prior-independent method, see \cite{Campeti:2022vom}):
\begin{itemize}
    \item the first consists of imposing a cut to the lowest values of the amplitude, typically setting the threshold value at some undetectable level for the considered experiments. For example, \citet{Cabass_2016} show the efficacy of this method while choosing $r>0.001$. Today, this threshold value is exactly the target sensitivity of experiments such as LiteBIRD; therefore, one may want to choose a less aggressive cut. Throughout the remainder of this paper, we will refer to this approach as the ``Single-Scale Approach'' (SSA).
    \item On the other hand, what \citet{Planck15_inflation} did to solve the issue (subsequently repeated in \citep{Planck_2018}) is to actually re-parameterize the tensor power spectrum using two different tensor-to-scalar ratios $(r_1,r_2)$ at two different pivot scales $(k_1,k_2)$, which are arbitrarily chosen. This, in fact, allows us to solve the problem: when $r_1$ approaches $0$, $r_2$ will do so accordingly without producing any pathological behavior. Then, from $(r_1,r_2)$ one can recover $(r_{\Tilde{k}},n_t)$ through 
    \begin{equation}
    \begin{aligned}
        n_t &= [\log \qty(r_2/r_1)/\log \qty(k_2/k_1)] + n_s - 1\ , \\
        r_{\Tilde{k}} &= r_1(\Tilde{k}/k_1)^{n_t - n_s +1}\ ,
    \label{eq: coordinate}
    \end{aligned}
    \end{equation}
    where $\Tilde{k}$ is some arbitrary scale (typically $\Tilde{k} = 0.01$ Mpc$^{-1}$). Specifically, \citet{Planck_2018} choose $(k_1,k_2) = (0.002, 0.02)$ Mpc$^{-1}$, imposing uniform priors on $(r_1,r_2)$. For obvious reasons, we will refer to this approach as the ``Two-Scales Approach'' (TSA).
\end{itemize}

\subsection{Robustness test: priors}\label{sec: test_prior}

Both approaches have their strengths and weaknesses; however, it is important to emphasize them to make a conscious choice on which to use. 

The first test we perform is to run an MCMC analysis on the priors alone, without introducing any other source of information (see Tab.\ref{tab: priors} for the specific values of the prior ranges), such as the likelihoods of the considered probes. In other words, this is a behavior check on the priors of the SSA and TSA, focused on the plane $(r_{0.01}, n_t)$ (also the $\Lambda$CDM parameters are kept free to vary). This allows us to obtain ``sampled'' priors to compare with the input ones and to gauge prior information on derived parameters. 
\begin{figure}[t]
    \centering
    \includegraphics[width = .45\hsize]{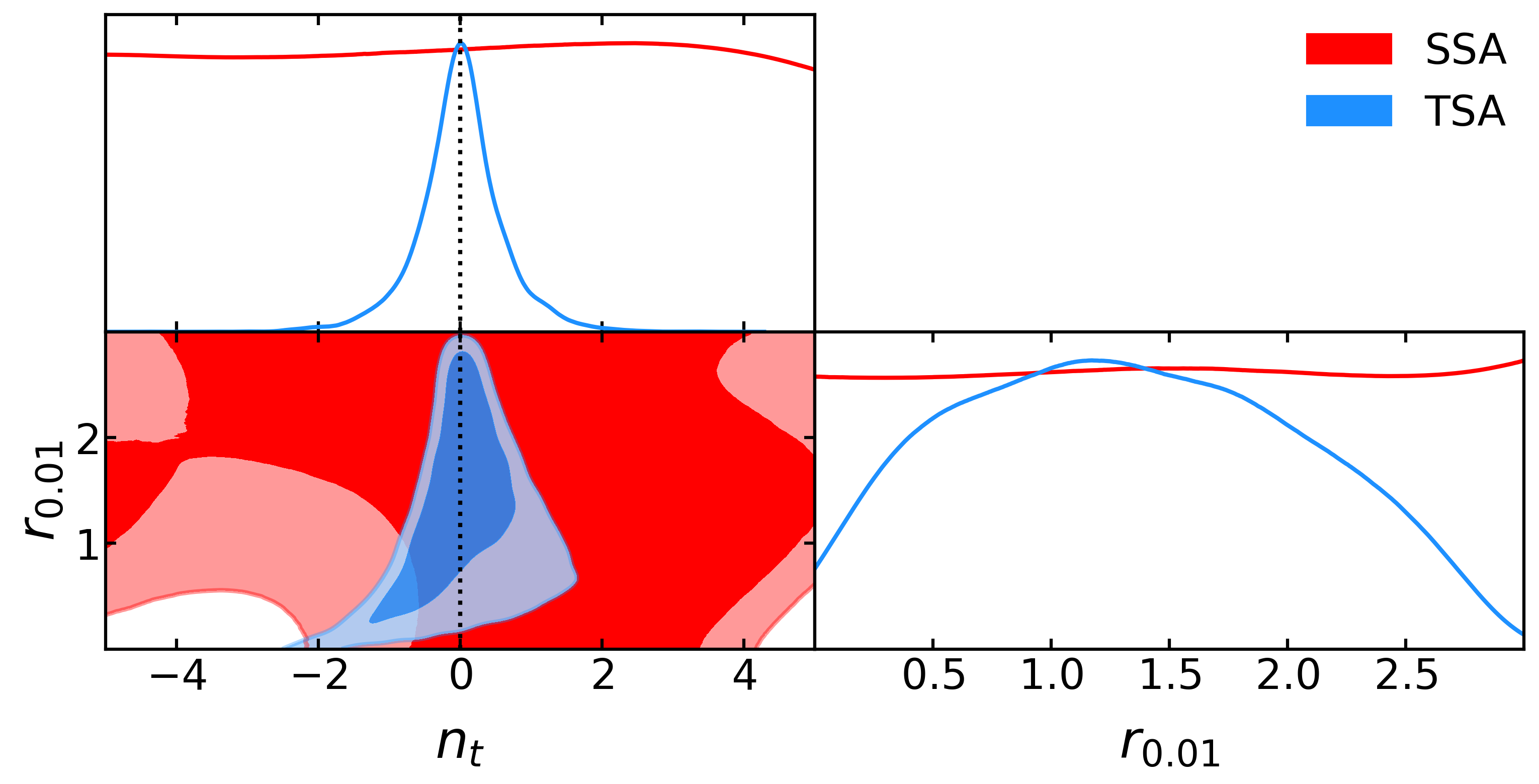}
    \caption{1D and 2D sampled priors on the $(r_{0.01}, n_t)$ plane, obtained using the SSA or the TSA. The vertical dotted line represents scale-invariance, i.e. $n_t = 0$.}
    \label{fig: priors}
\end{figure}

\begin{figure}[t]
    \centering
    \includegraphics[width = .49\textwidth]{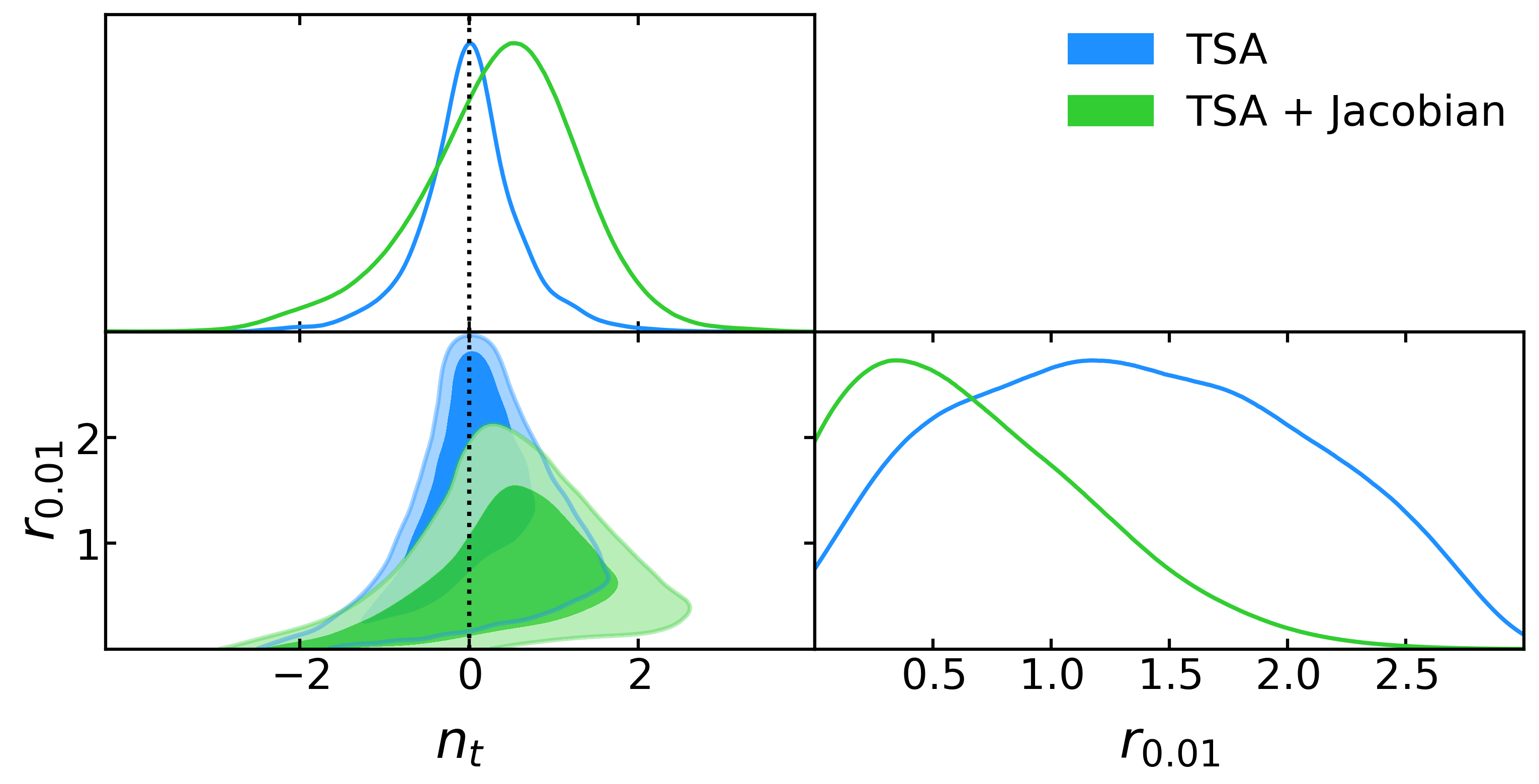}
    \hfill
    \includegraphics[width = .49\textwidth]{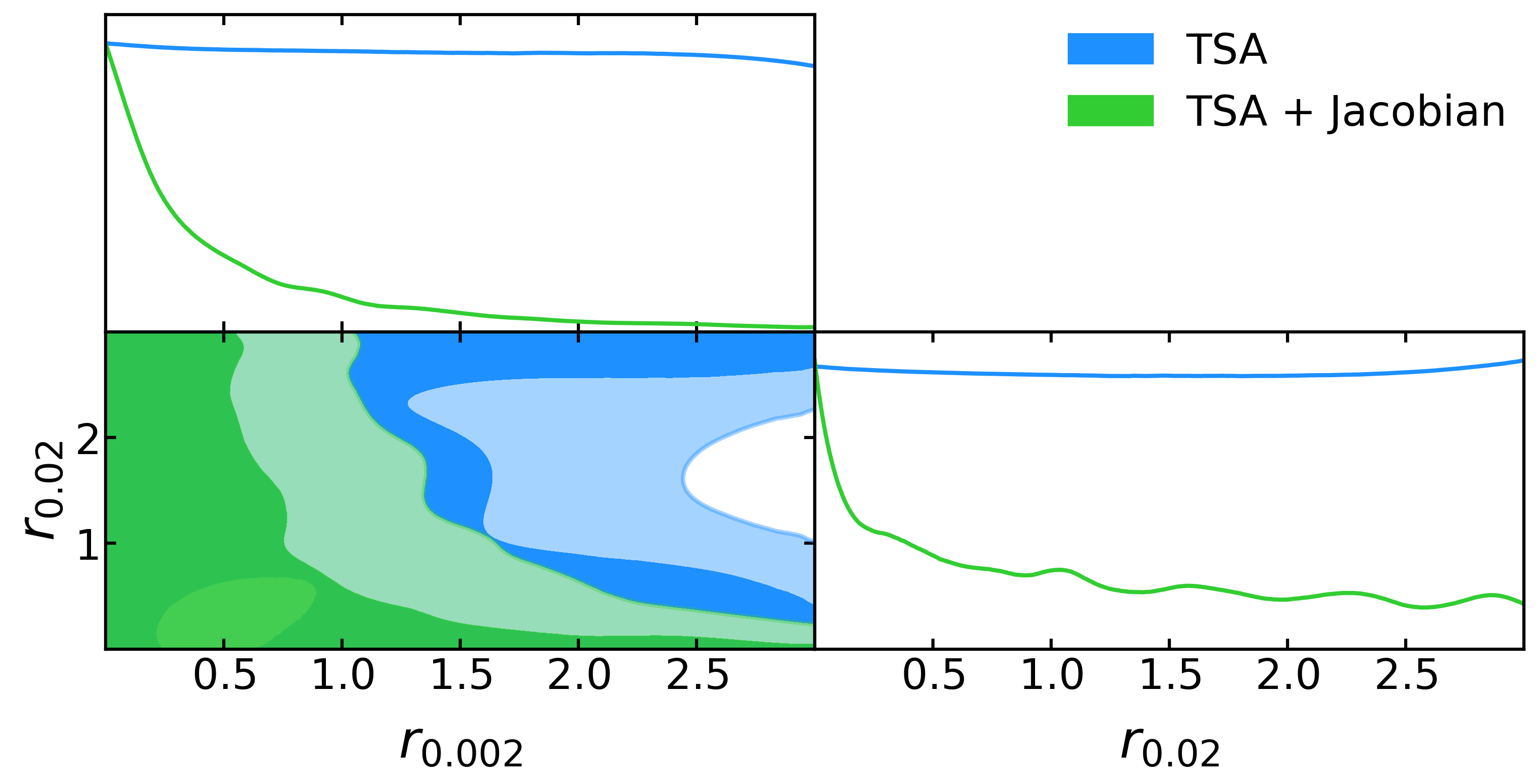}
    \caption{1D and 2D sampled priors on the $(r_{0.01}, n_t)$ plane (left) and on the $(r_1, r_2)$ plane (right), obtained using the TSA with and without the Jacobian reweighting (see eq.\ref{eq: jacobian}). The vertical dotted line represents scale-invariance, i.e. $n_t = 0$.}
    \label{fig: jacobian_scales}
\end{figure}

Figure \ref{fig: priors} shows the marginalized 1D distributions and the marginalized 2D 68\% and 95\% CL\footnote{From now on, every 2D plot will show the 68\% and 95\% CL.}. Here, we can see that the sampled prior on $(r_{0.01}, n_t)$ using the SSA are flat, as expected since the sampling is performed directly on those two parameters. For what regards the TSA, figure \ref{fig: priors} instead shows a neat preference for values of $r_{0.01}$ different from zero, thus mimicking a detection (for the derived parameter $r_{0.01}$ here we impose $r_{0.01} \in [0,3]$). The marginalized distribution gives $0.08<r_{0.01}<2.61$ at $95\%$ CL. In addition, there is a pretty strong pull on $n_t$ towards scale invariance ($n_t = 0$), giving $-1.33<n_{t}<1.31$ at $95\%$ CL. This suggests that, while this method will solve the pathological behavior of $n_t$, it might also introduce a bias, especially if the dataset does not well constrain these parameters (in this case, the likelihood will dominate the final posterior). In order to explore this behavior more thoroughly, we also repeat this latter test on the TSA while reweighting the sample on the Jacobian of the transformation $\qty(r_1,r_2) \leftrightarrow \qty(r_{\Tilde{k}},n_t)$:
\begin{equation}
    J = \frac{r_{0.01}}{r_1 r_2} \times \frac{1}{\log\qty(k_1/k_2)} \ .
    \label{eq: jacobian}
\end{equation}
Indeed, this procedure has been used in the literature to alleviate this bias introduced by the prior (e.g. \textit{Planck 2018}). As the left panel of figure \ref{fig: jacobian_scales} shows, this seems to mitigate the amount of information introduced by the priors of the TSA but does not completely solve the problem. 
\begin{figure}[t]
    \centering
    \includegraphics[width = .49\textwidth]{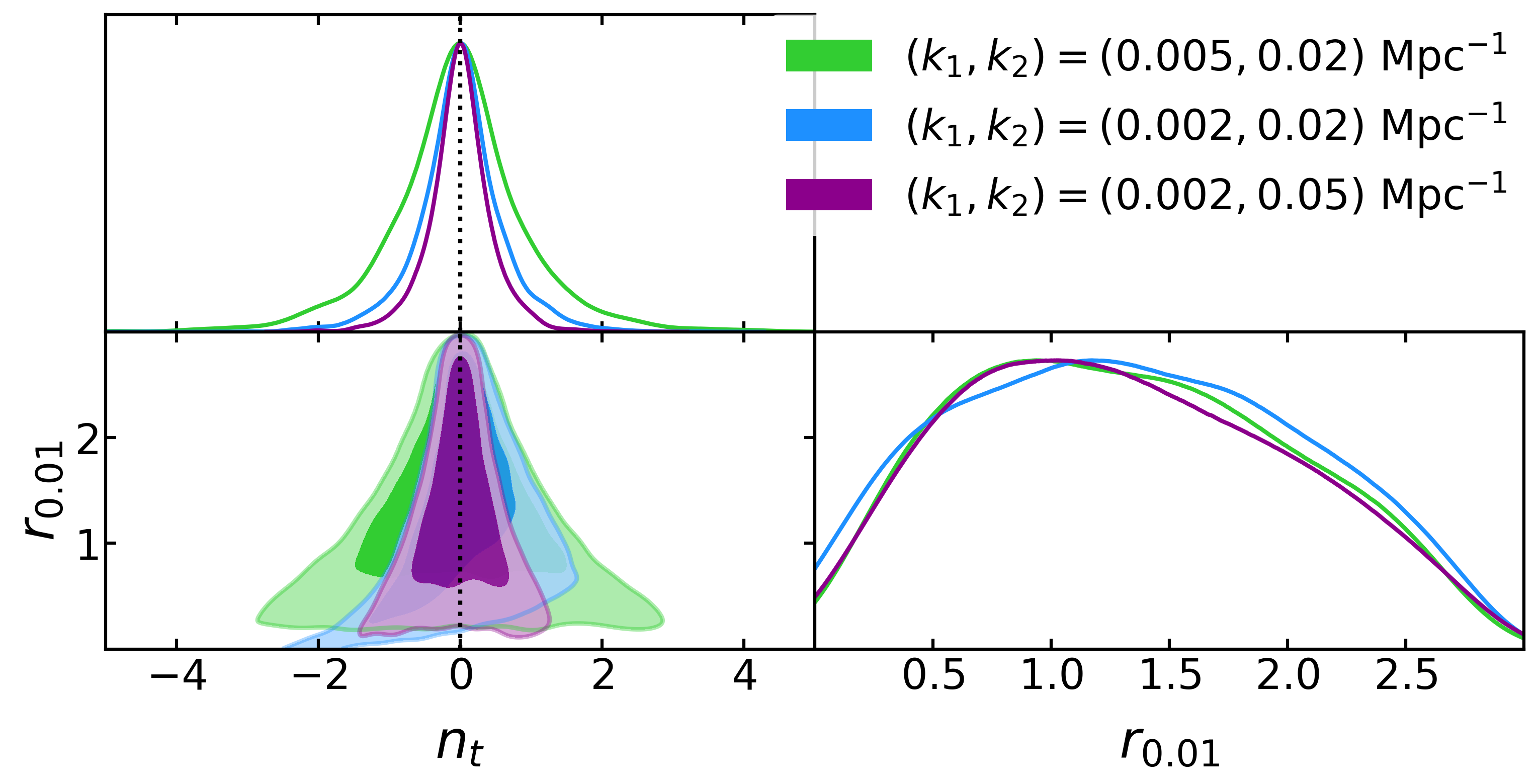}
    \caption{1D and 2D sampled priors on the $(r_{0.01}, n_t)$ plane, obtained using the TSA with either $(k_1,k_2) = (0.005,0.02), (0.002,0.02), (0.002,0.05)$ Mpc$^{-1}$. The vertical dotted line represents scale-invariance, i.e. $n_t = 0$.}
    \label{fig: jacobian_scales2}
\end{figure}
In particular, we obtain $r_{0.01} < 1.67$ and $-1.90 < n_t < 2.11$ at $95\%$ CL. Thus, on one side the sampled priors are not mimicking a detection of the tensor-to-scalar ratio anymore, however, we still obtain a range of variation of $n_t$ roughly equal to 4 units. This same range is very similar to what we obtain using the SSA on the state-of-the-art dataset (see section \ref{sec: Results}), thus the results obtained through the TSA might be affected by bias, even when including the Jacobian. The right panel on figure \ref{fig: jacobian_scales} shows that the sampled priors on $r_1$ and $r_2$ are correctly flat and are remapped to non-flat distributions by the Jacobian. In accordance with what we show here, for example, \textit{Planck} 2018 shows that re-weighting the sampling allows obtaining a final posterior shifted towards $r_{0.01}=0$ \cite{Planck_2018}.

In the TSA, we can further test the choice of arbitrary scales $(k_1,k_2)$, which in \citep{Planck_2018} are chosen to be $(0.002,0.02)$ Mpc$^{-1}$. Thus, we repeat the prior analysis for $(k_1,k_2) = (0.002,0.05)$ Mpc$^{-1}$ and $(k_1,k_2) = (0.005,0.02)$ Mpc$^{-1}$. Notice that in the former case, we increase the separation between $k_1$ and $k_2$, while in the latter case, we decrease it. In other words, we are testing the dependence of the prior on the leverage arm given by $k_2 - k_1$, which will affect the capability to recover $n_t$. 
Figure \ref{fig: jacobian_scales2} shows that the sampled prior in the tensor-to-scalar ratio is only partially affected by the choice of scales, while the one in the tilt changes significantly as expected. Indeed, a larger leverage arm exacerbates the preference for scale invariance and vice versa. Thus, while using the TSA, one must be careful of what scales to use. For example, the largest and smallest scales that an experiment has access to should provide an upper bound on the leverage arm length. However, going for this choice will produce the most peaked prior distribution possible on $n_t$.

Despite this, both cases analyzed here seem to break more prominently the degeneracy between $r_{0.01}$ and $n_t$ w.r.t. the choice $(k_1,k_2) = (0.002,0.02)$ Mpc$^{-1}$. Notice that this test depends to some degree on the fact that the MCMC sample is finite. Indeed, after convergence, the MCMC stops exploring the parameter space, potentially under-representing low-probability regions. In appendix \ref{sec: coords} we perform it in an alternative way, i.e. directly from the coordinate transformation equations shown above, and so taking care of this effect. The conclusions do not change following this method, however, we can explain better the behavior of the region at low $r_{0.01}$ and high $n_t$ of figure \ref{fig: jacobian_scales} (left panel). Indeed, in that low-probability region, there are no samples to be re-weighted, thus the probability drops to zero because the sample is finite (see figure \ref{fig: analytical_priors} in appendix \ref{sec: coords}). In this specific case, this leads to an underestimation of the upper bound of the tilt. In general, this may or may not be the case depending on the specific choice between $k_1$ and $k_2$. Thus, together with the leverage arm, one should also keep this in mind when choosing those scales. 

\subsection{Robustness test: mock dataset}\label{sec: test_exact}

We perform another robustness check on the two approaches: we fix the 6 $\Lambda$CDM parameters to the best-fit values of \textit{Planck}(2018) \cite{Planck_parameters} and we allow only the tensor sector to vary. Then, instead of running the MCMC analysis on actual data (i.e. with the likelihood of some experiment), we define an exact likelihood for the B-mode spectrum \cite{gerbino_likelihood_2020}:
\begin{equation}
    -2\log \mathcal{L} = \sum_\ell (2\ell+1)\qty[\frac{\hat{C}_\ell^{BB}}{\bar{C}_\ell^{BB}} - \log\qty|\frac{\hat{C}_\ell^{BB}}{\bar{C}_\ell^{BB}}| - 1] \ .
\end{equation}

Here, $\hat{C}_\ell^{BB}$ is a fiducial BB spectrum, and $\bar{C}_\ell^{BB}$ is the theoretical prediction based on the MCMC step. Notice, that we do not introduce any source of instrumental noise. Instead, we use lensing as our noise level: indeed, we can write both the fiducial and theoretical spectra as $C_\ell^{BB} = C_\ell^{prim.} + C_\ell^{lens.}$, where $C_\ell^{prim.}$ is the primordial spectrum coming from the presence of an inflationary background of GWs, obtained fixing $r_{0.01}, n_t$, whereas $C_\ell^{lens.}$ is the contribution given by the lensing of polarized photons by mass distributions (see figure \ref{fig: exp}). In other words, this likelihood is representing a mock dataset from which we try to extract the fiducial values of the tensor parameters and their 95\% CL intervals.

\begin{figure}[t]
    \centering
    \includegraphics[width = .49\textwidth]{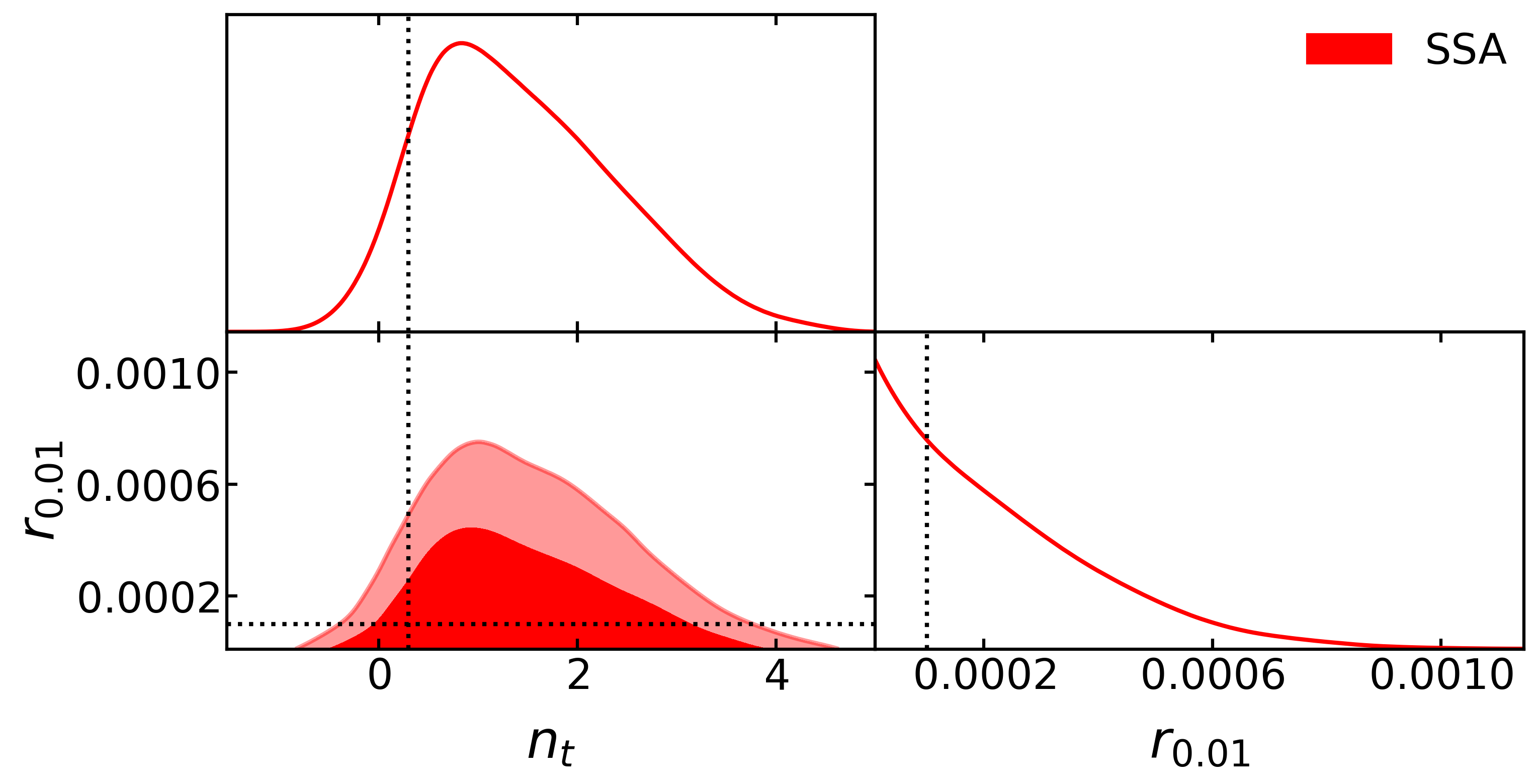}
    \includegraphics[width = .49\textwidth]{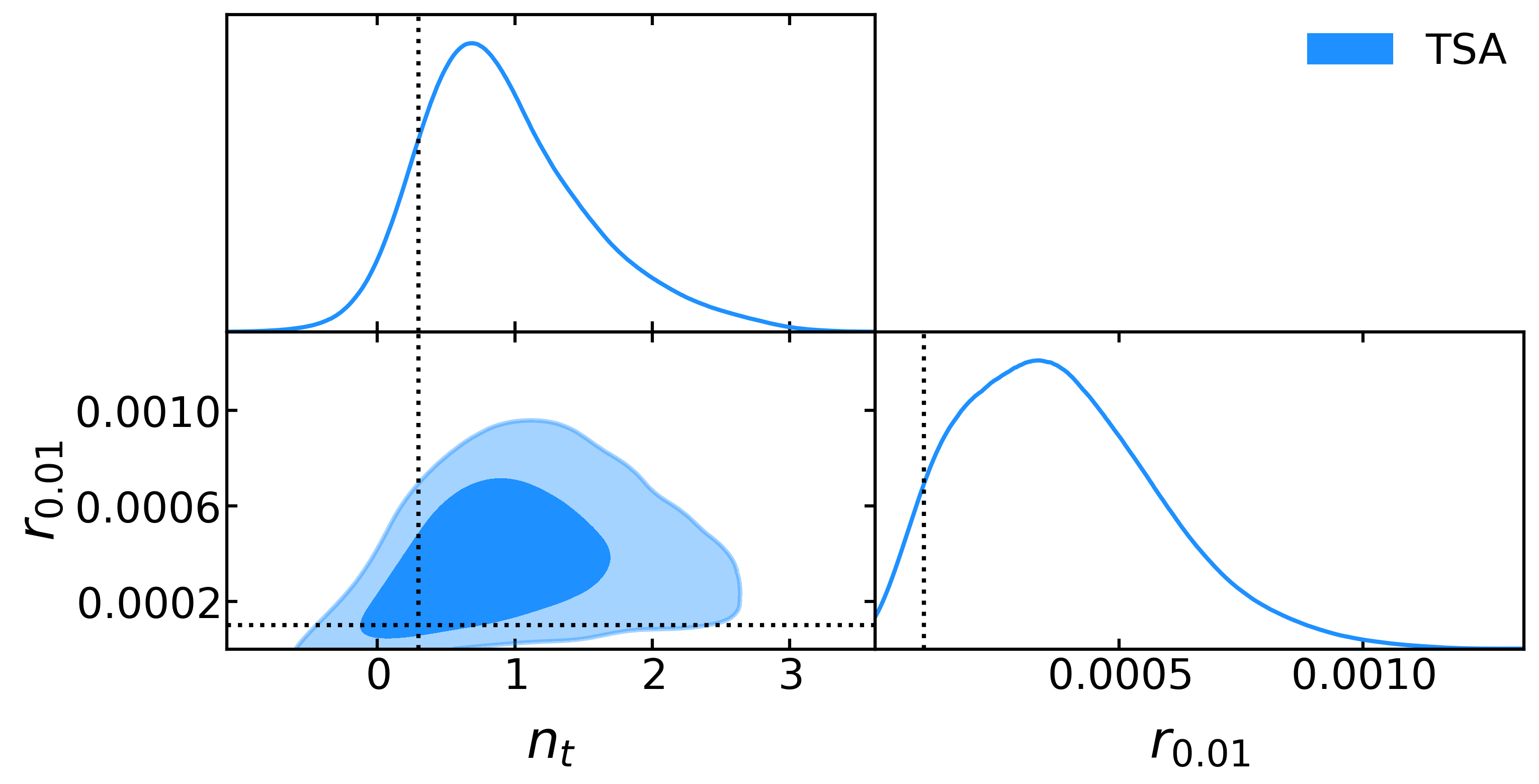}
    \caption{1D and 2D posteriors on the $(r_{0.01}, n_t)$ plane, obtained using the SSA with an exact likelihood on the B-mode spectrum. The dotted markers indicate the fiducial values for $(r_{0.01}, n_t) = (10^{-4}, 0.3)$.}
    \label{fig: exact}
\end{figure}

As for the fiducial choice of the BB spectrum, we want to reproduce the case in which the primordial spectrum is below the noise, in order to see how the two approaches deal with a non-detection of the tensor spectrum. For this reason, we choose $(r_{0.01}, n_t) = (10^{-4}, 0.3)$. The results are summarized by the left and right panels of figure \ref{fig: exact}.
The 95\% CL ranges that we obtain are instead shown in table \ref{tab: exact_ranges}.
\begin{table}[t]
 \centering
\begin{tabular}{ccc}
    \toprule
    $\mathbf{r_{0.01}}$ & \textbf{Best-fit} & \textbf{95\% CL} \\ \midrule
    SSA & $1.00 \times 10^{-4}$ & ${[}0.1, 6.0{]}\times 10^{-4}$ \\
    TSA & $0.95 \times 10^{-4}$ & ${[}0.2, 7.6{]}\times 10^{-4}$ \\ \midrule
    $\mathbf{n_t}$ & \textbf{Best-fit} & \textbf{95\% CL} \\ \midrule
    SSA & $0.30$ & ${[}-0.24, 3.43{]}$ \\
    TSA & $0.30$ & ${[}-0.15, 2.30{]}$ \\ \bottomrule
\end{tabular}
\caption{$95\%$ CL and best-fit values obtained with SSA and TSA when using an exact likelihood on the B-mode spectrum. The fiducial values are $(r_{0.01}, n_t) = (10^{-4}, 0.3)$.}
  \label{tab: exact_ranges}
\end{table}

\begin{figure}[t]
    \centering
    \includegraphics[width = .49\textwidth]{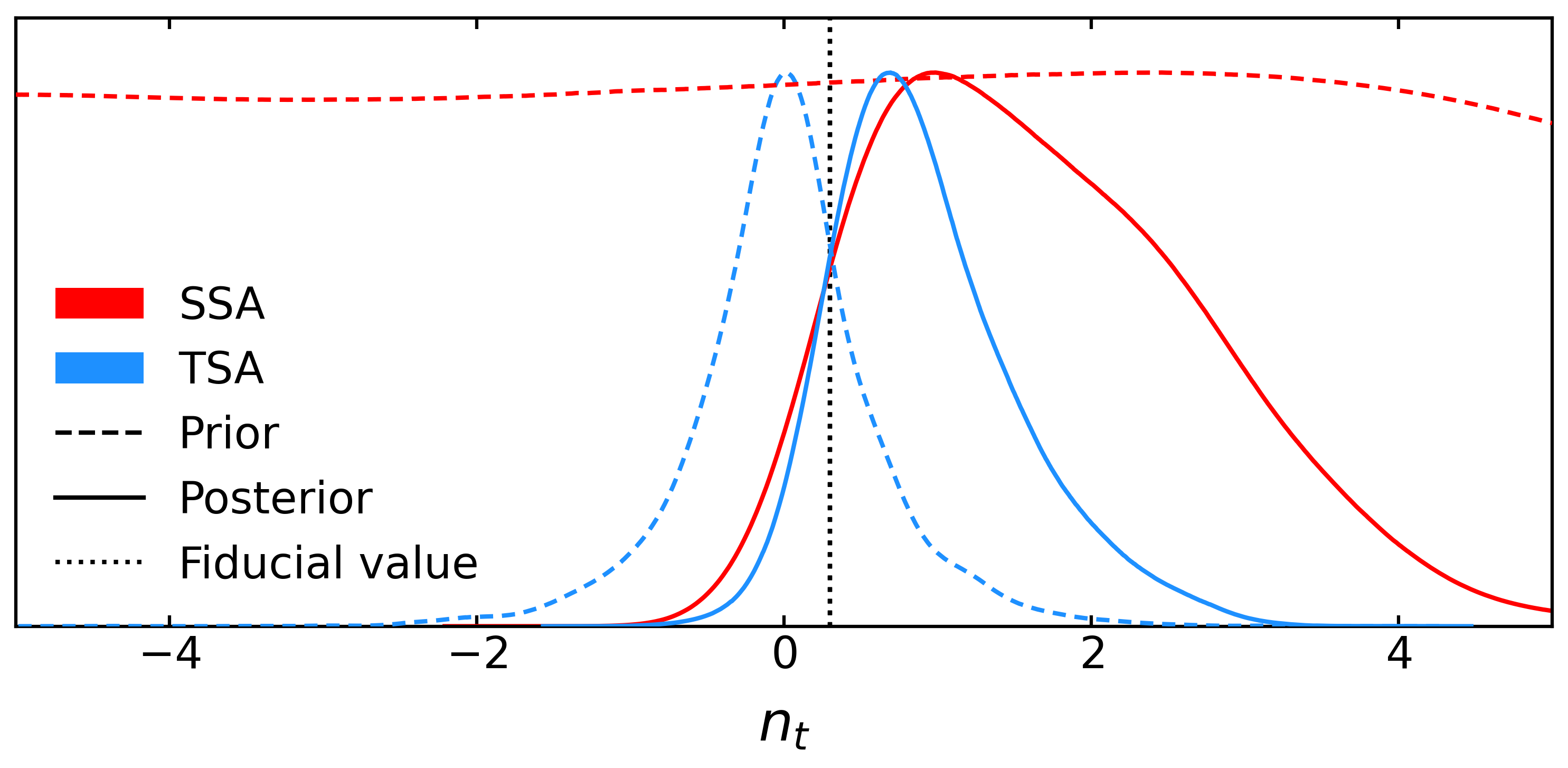}
    \caption{1D posterior on the $n_t$, obtained using the SSA and the TSA with an exact likelihood on the B-mode spectrum. The dashed lines represent the sampled prior distributions obtained in figure \ref{fig: priors}. The dotted line indicates the fiducial value of $n_t=0.3$.}
    \label{fig: comparison_exact}
\end{figure}

From this table, together with the figure, one can see that both approaches are able to recover approximately the correct value of the amplitude and tilt as the best fit of the MCMC run. With regards to the estimation of the $95\%$ CL, the SSA provides a slightly stricter bound on the amplitude and a broader one on the tilt, as one can expect from section \ref{sec: test_prior}. 
Figure \ref{fig: comparison_exact} highlights this by comparing the posterior distributions in $n_t$ obtained with the exact likelihood with the priors shown in figure \ref{fig: priors}. Notice that the distributions of figure \ref{fig: comparison_exact} are normalized to their maximum.
We can also test an arbitrary choice affecting the SSA, i.e. the cut-off at low $r$. In fact, SSA results could be driven by marginalization effects: in a multidimensional problem, if a large part of the probability volume is in a certain area, the final posterior will be drawn toward that region just as a result of the marginalization procedure that one performs when giving the result on a single parameter \cite{gomez-valent2022FastTestAssessimpact}. 

To investigate this potential problem, as a first step, we repeat this analysis assuming $r_{0.01}^{cut} = 10^{-4}, 10^{-5}, 10^{-6}$. Notice that the first value is equal to the fiducial tensor-to-scalar ratio, so that case will show what happens if we cut some relevant part of the parameter space, marginalizing the rest. The other two cutoffs are below the fiducial value; thus, they will test the stability of the estimate when we exclude a region with little-to-no posterior volume.
\begin{figure}[t]
    \centering
    \includegraphics[width = .5\textwidth]{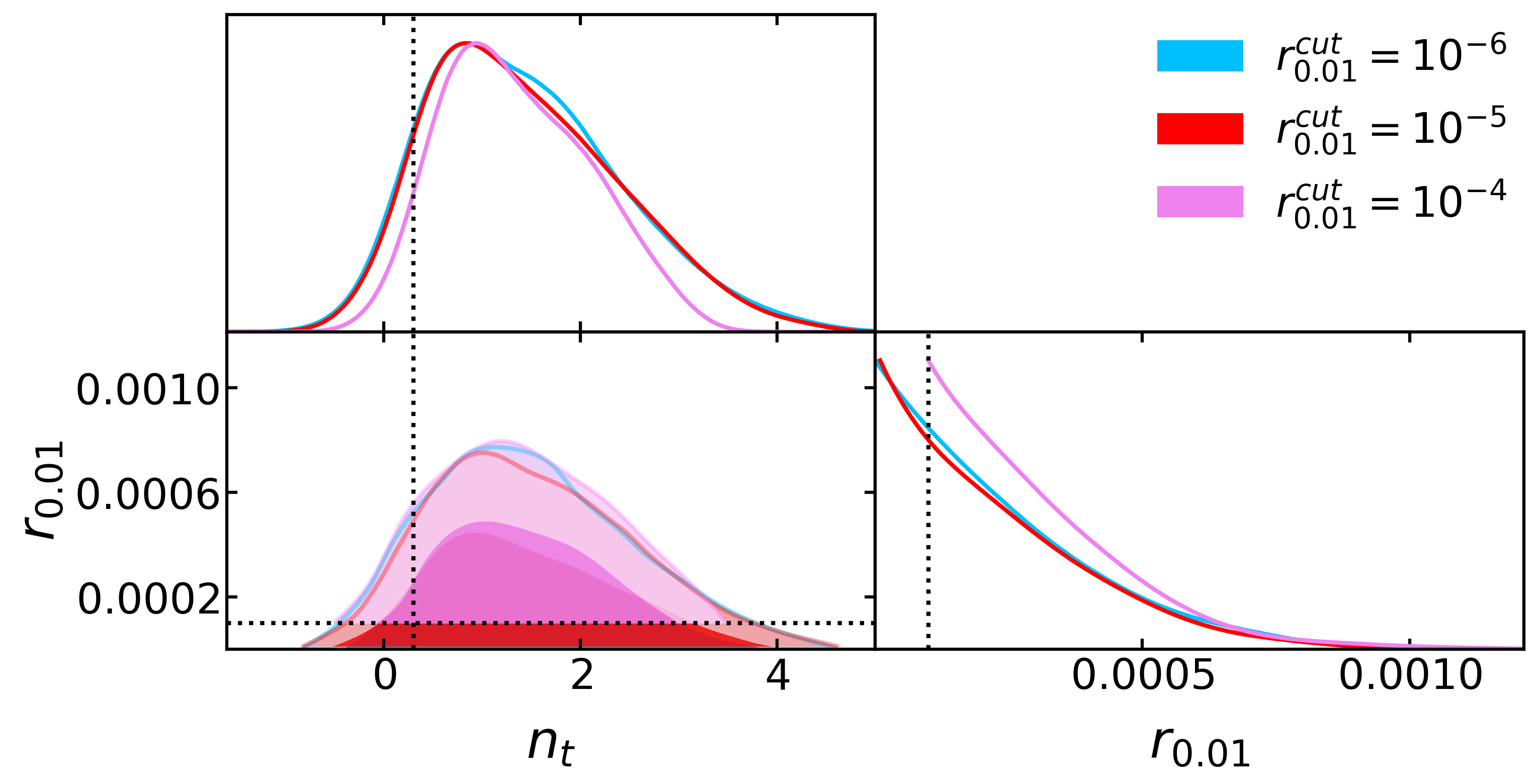}
    \caption{1D and 2D posteriors on the $(r_{0.01}, n_t)$ plane, obtained using the SSA and changing the cut-off at low amplitudes: $r_{0.01}^{cut} = 10^{-4}, 10^{-5}, 10^{-6}$. The dotted markers indicate the fiducial values for $(r_{0.01}, n_t) = (10^{-4}, 0.3)$.}
    \label{fig: cutoffs}
\end{figure}
Focusing on the first cutoff value, figure \ref{fig: cutoffs} shows that cutting an interesting part of the parameter space does have consequences on the final posterior, as expected. Indeed, looking at the 1D posterior of $r_{0.01}$ the posterior area below $r_{0.01}^{cut} = 10^{-4}$ is redistributed on the allowed region, providing a worse bound on the parameter.
On the other hand, the estimate of $n_t$ improves. In fact, the most extreme values of the tilt are obtained by the MCMC analysis for very low values of $r_{0.01}$, which in this case are completely neglected.

Then, focusing on the other two cases, figure \ref{fig: cutoffs} shows that the analysis is stable in both $r_{0.01}, n_t$ as soon as the cut is performed underneath the fiducial value since the corresponding posterior volume is small. Indeed, the posteriors obtained in these cases are nearly identical. Table \ref{tab: cutoffs} summarizes the results of this test.
\begin{table}[t]
\centering
\begin{tabular}{ccc}
    \toprule
    $\mathbf{r_{0.01}}$ & \textbf{Best-fit} & \textbf{95\% CL} \\ \midrule
    $r_{0.01}^{cut} = 10^{-6}$ & $1.03 \times 10^{-4}$ & ${[}0.01, 6.09{]}\times 10^{-4}$ \\
    $r_{0.01}^{cut} = 10^{-5}$ & $1.00 \times 10^{-4}$ & ${[}0.10, 5.95{]}\times 10^{-4}$ \\
    $r_{0.01}^{cut} = 10^{-4}$ & $1.11 \times 10^{-4}$ & ${[}1.00, 6.25{]}\times 10^{-4}$ \\ \midrule
    $\mathbf{n_t}$ & \textbf{Best-fit} & \textbf{95\% CL} \\ \midrule
    $r_{0.01}^{cut} = 10^{-6}$ & $0.30$ & ${[}-0.28, 3.54{]}$ \\
    $r_{0.01}^{cut} = 10^{-5}$ & $0.30$ & ${[}-0.24, 3.43{]}$ \\
    $r_{0.01}^{cut} = 10^{-4}$ & $0.31$ & $\ {[}0.002, 2.92{]}$ \\ \bottomrule
\end{tabular}
\caption{$95\%$ CL and best-fit values obtained with SSA when using an exact likelihood in the B-mode spectrum. The fiducial values are $(r_{0.01}, n_t) = (10^{-4}, 0.3)$. Here, we also assumed a different value of the cutoff at low amplitudes, $r_{0.01}^{cut} = 10^{-4}, 10^{-5}, 10^{-6}$.}
\label{tab: cutoffs}
\end{table}

\begin{figure}[t]
    \centering
    \includegraphics[width = .5\textwidth]{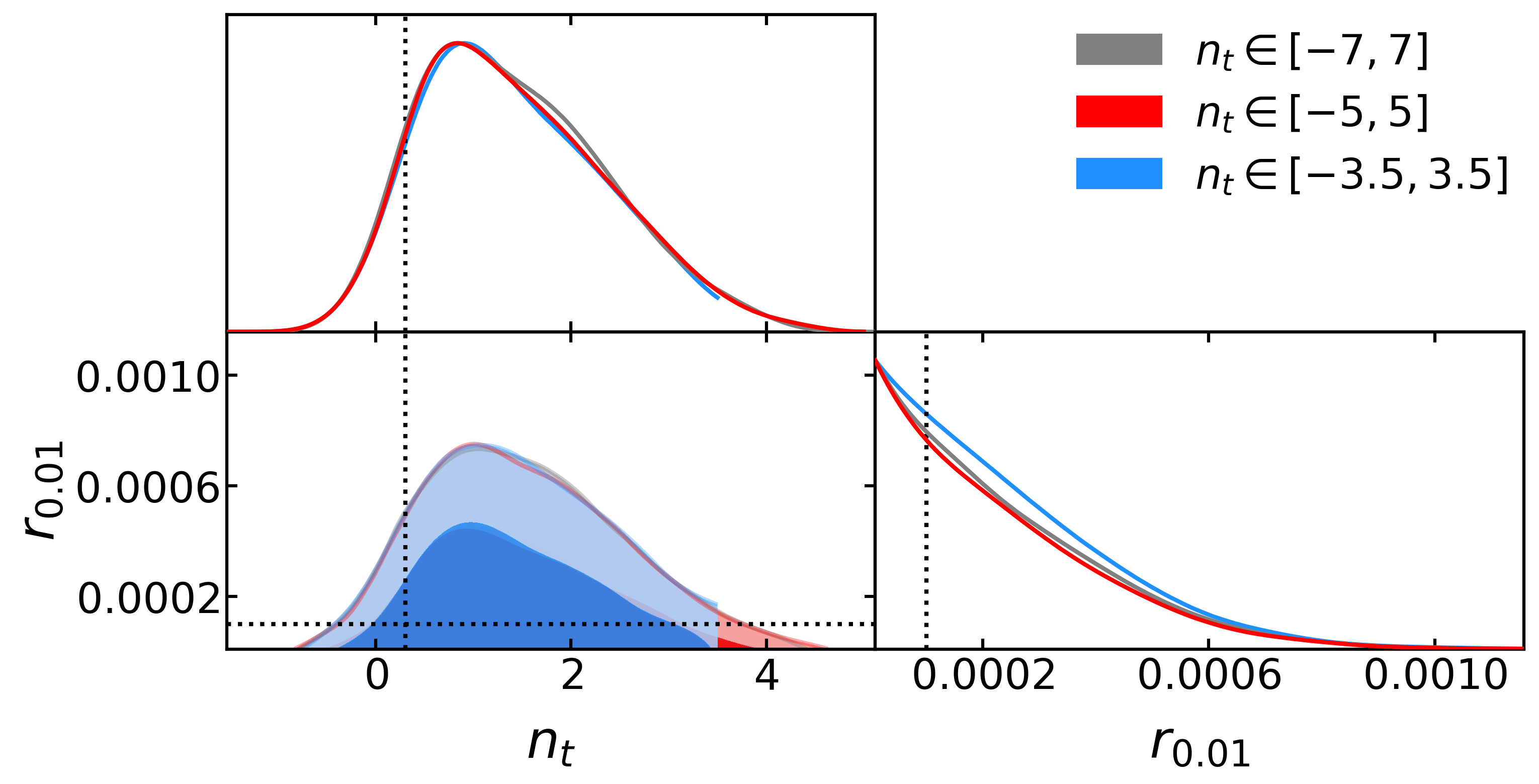}
    \caption{1D and 2D posteriors on the $(r_{0.01}, n_t)$ plane, obtained using the SSA while changing prior range on $n_t$: $n_t \in \qty[-5,5]$, $n_t \in \qty[-7,7]$ and $n_t \in \qty[-3.5, 3.5]$.}
    \label{fig: nt_priors}
\end{figure}

Secondly, we inquire about an eventual over-weighting of the region near the cut-off in $r_{0.01}$ when using the SSA. Indeed, we already said that without the cutoff $n_t$ would be completely unconstrained near $r=0$. This means that near $0$ there would be a lot of posterior volume, producing an over-weighting of the region of low tensor-to-scalar ratio, affecting the final results through the marginalization procedure. If this is the case, the contours should be dependent on the width of the $n_t$-prior, since the larger it is, the more available volume there would be. The prior is changed from $n_t \in \qty[-5,5]$ to $n_t \in \qty[-7,7]$ and to $n_t \in \qty[-3.5, 3.5]$ (cutting a relevant part of the posterior volume). The results are shown in figure \ref{fig: nt_priors}, and, as expected, the difference comes from how much posterior volume has been excluded in the various cases.

Therefore, once the cutoff $r_{0.01}^{cut}$ (prior on $n_t$) is taken sufficiently low (large enough) to avoid excluding a relevant portion of the posterior volume, the SSA results will not change significantly.

We mentioned that SSA and TSA are two Bayesian ways to solve the non-detection issue when we estimate tensor parameters; however, we can also use an alternative prior-independent frequentist method to shed some light on the results we obtain. In fact, figure \ref{fig: exact} already shows that the 1D marginalized distribution on $r_{0.01}$ increases towards zero. This method is called the Profile Likelihood (PL) technique \cite{planck_int_2014_pl, Herold:2021ksg, Campeti:2022vom, gomez-valent2022FastTestAssessimpact}. Without entering the details, the underlying idea of this frequentist technique is the following: consider a multidimensional parameter estimation, where the parameters are $\theta_1, \theta_2, \dots, \theta_N$. If $\mathcal{P}(\theta_1, \theta_2, \dots, \theta_N)$ is the probability distribution of your parameters, you can calculate the PL of $\theta_1$ by maximizing $\mathcal{P}$ on all other parameters while keeping $\theta_1$ fixed. In fact, repeating the process on a list of values of $\theta_1$ can map the probability in the region considered. Note that no integration is performed; thus, this is not subject to volume effects. Instead, all other parameters are ``conspiring'' to maximize the probability, in order to provide the region of $\theta_1$ that fits well the data. 
So, we perform a PL on this mock dataset and we compare the marginalized results of both SSA and TSA. Note that this test tells us how much the SSA or TSA results are affected by marginalization effects.
\begin{figure}[t]
    \centering
    \includegraphics[width = .49\textwidth]{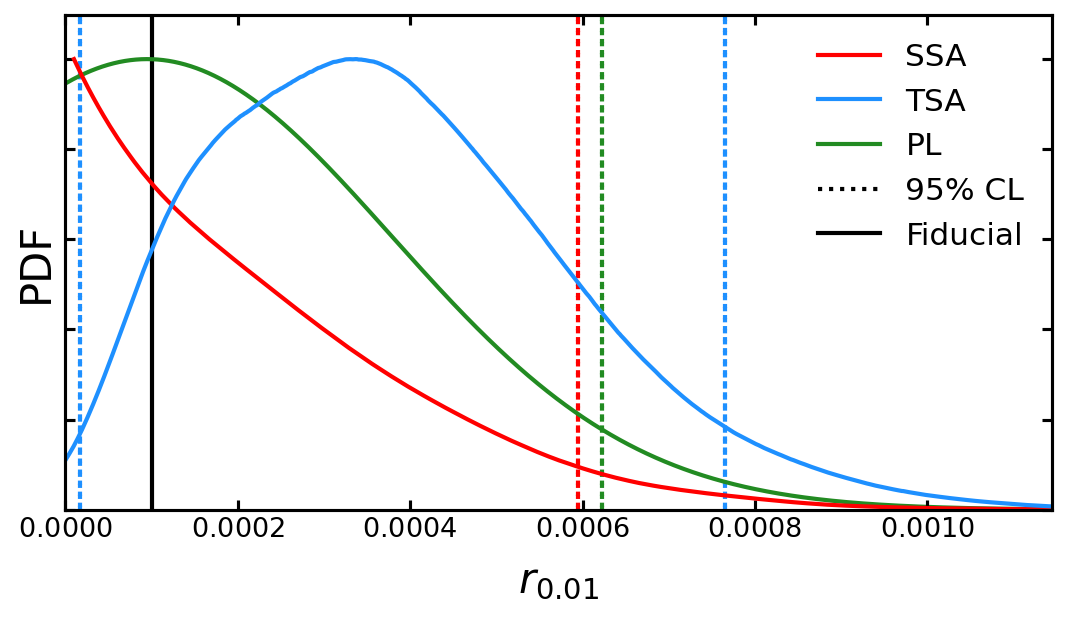}
    \includegraphics[width = .49\textwidth]{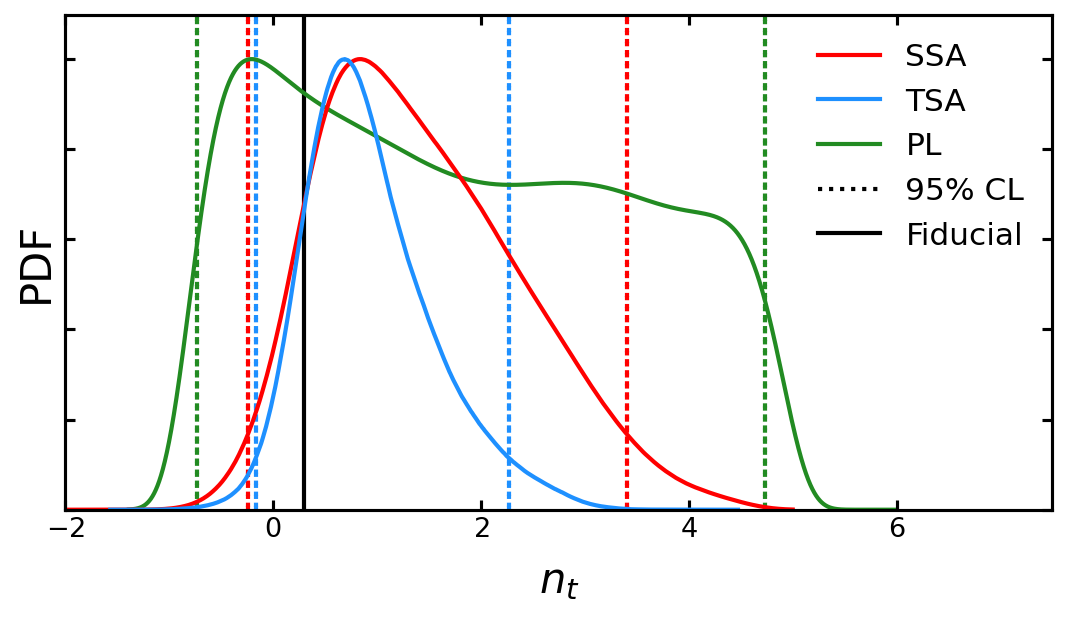}
    \caption{1D results on $r_{0.01}$ and $n_t$, obtained using SSA, TSA, and PL with an exact likelihood on the B-mode spectrum. The black solid lines report the fiducial values of  $r_{0.01}$ and $n_t$. The vertical dotted lines represent the boundaries of the SSA and TSA 95\% CL intervals. Note that SSA gives a single-tail bound on $r_{0.01}$ while the TSA provides a two-tail one. Also, the green dotted lines indicate the 95\% credible intervals of the PL. These are also summarized in table \ref{tab: MCMC_vs_PL}.}
    \label{fig: PLvsMCMC}
\end{figure}
\begin{table}[t]
\centering
\begin{tabular}{ccc}
    \toprule
     & $\mathbf{r_{0.01}}$ \textbf{95\% CL} & $\mathbf{n_t}$ \textbf{95\% CL} \\ 
    \midrule
    SSA & $< 6.0 \times 10^{-4}$ & ${[}-0.24,\ 3.43{]}$ \\
    TSA & ${[}0.2,\ 7.6{]} \times 10^{-4}$ & ${[}-0.15,\ 2.30{]}$ \\
    \midrule
    & $\mathbf{r_{0.01}}$ \textbf{95\% credible interval} & $\mathbf{n_t}$ \textbf{95\% credible interval} \\ 
    \midrule
    PL  & $< 6.2 \times 10^{-4}$ & ${[}-0.73,\ 4.74{]}$ \\ 
    \bottomrule
\end{tabular}
\caption{$95\%$ CL obtained with SSA and the TSA compared to the 95\% credible intervals of the PL.}
\label{tab: MCMC_vs_PL}
\end{table}
Figure \ref{fig: PLvsMCMC} shows the distributions obtained, together with the $95\%$ CL intervals (see table \ref{tab: MCMC_vs_PL} for the actual results), which for the PL case are obtained by integration of the curve up to the value (or values in the case of $n_t$) where it gives the $95\%$ of the total area \cite{gomez-valent2022FastTestAssessimpact}. 
Starting from the left panel of figure \ref{fig: PLvsMCMC}, we note that PL is completely compatible with zero, as expected, and the $95\%$ credible interval excludes the region above $r_{0.01}>6.2\times 10^{-4}$. Comparing this with the SSA, we can see that in fact the 95\% CL upper bound on $r_{0.01}$ is slightly smaller, precisely of $3.2\%$ w.r.t. the PL one. This is due to the fact that the marginalized volume increases toward $r_{0.01}=0$. In spite of this, the results are quite consistent with the PL bound, suggesting that the marginalization effect may not be a major concern for the SSA. Looking at the TSA instead, it is clear that the prior volume in $r_{0.01}$ pushes the results to higher values. In fact, a large area below the blue curve is above the 95\% limit of the PL, causing a difference in the bound obtained of $22.6\%$ (comparing the PL and the TSA upper bounds). Also, this confirms that the presence of a lower bound, i.e. the exclusion of $r_{0.01}=0$, is totally drawn by the prior volume, as expected from section \ref{sec: test_prior}. 

Focusing now on the right panel of figure \ref{fig: PLvsMCMC}, the PL gives a $95\%$ credible interval of $-0.73<n_t<4.74$, presenting a probability peak around $n_t\simeq -0.21$. This time, the comparison with SSA and TSA shows that both methods are affected by marginalization effects. Despite this, the SSA is the most conservative of the two, since it is not affected by the large prior volume around scale invariance characterizing the TSA.

In conclusion, we recall that the PL is a complementary tool to a Bayesian analysis. Together, they allow us to get a more complete picture of the parameter space we are exploring. For example, a measurement of $n_t=4.7$ would seem in tension with our SSA result (and even more with the TSA one). Still, PL tells us that the model fits well data there, being inside the $95\%$ credible interval. 
Vice versa, a measurement of $r_{0.01}=7.6 \times 10^{-4}$ is plausible according to the TSA but is excluded by the PL at the $98.5\%$ level. In other words, now we can gauge the marginalization effects on the two Bayesian methods. Thus, we will keep this in mind for the following sections.

In light of these tests and given that we want to provide reliable bounds in the tensor sector, we choose to produce our baseline results with the SSA, assuming $r_{0.01}^{cut} = 10^{-5}$. In fact, one of the most promising models for describing tensor modes is the Starobinsky model \cite{Starobinsky}, which predicts $r \approx 0.004$, so $10^{-5}$ is more than two orders of magnitude lower, as was $10^{-6}$ in the robustness test. For $n_t$ we stick to the prior range reported in table \ref{tab: priors}. These choices ensure that we do not exclude a large portion of the posterior volume, thus we do not introduce any additional marginalization effect. This would artificially increase the upper bound of $r_{0.01}$ as a result of having used an aggressive $r_{0.01}^{cut}$, or having excluded a priori some inflationary model characterized by a high $\abs{n_t}$. 

In spite of this, in appendix \ref{sec: TSA} we report the results obtained with the TSA, while assuming $(k_1,k_2) = (0.002,0.02)$ Mpc$^{-1}$ (as done in \cite{Planck_2018}) and no Jacobian transformation re-weighting.

\section{Datasets} \label{sec: data}

Before going to the new bounds on $\qty(r_{0.01},n_t)$, we must understand what data are available in the market. We will divide this section according to the different experiments. Here, we just collect the ones that have been used to obtain the current bounds on the tensor parameters and those we want to use to update the constraints. As partially shown in figure \ref{fig: exp}, there are many others that we will not mention.

\subsection{\textit{Planck} satellite}

In most parameter estimation problems concerning CMB, \textit{Planck} satellite data play a key role. Indeed, the current bounds on $\qty(r_{0.01},n_t)$ are obtained using data from \textit{Planck} Release 3 (PR3) in the form of publicly available likelihoods (see \cite{Planck_like} and \cite{Ade_2014, Ade_2018} for further details). In particular, sticking to the common nomenclature, we will call the ones used here:
\begin{itemize}
    \item ``\textit{plikTTTEEE}'', encoding the high-$\ell$ parts of CMB temperature, E-mode polarization, and their cross-correlation, i.e. the TT, EE, and TE spectra \citep{Planck_like};
    \item ``\textit{lowlTT}'', encoding the low-$\ell$ parts of the TT spectrum \citep{Planck_like};
    \item ``\textit{lowlEE}'', encoding the low-$\ell$ parts of the EE spectrum \citep{Planck_like};
    \item ``\textit{lensing}'': encoding the presence of gravitational potentials along the line of sight, which will convert EE $\to$ BB \citep{Planck_like, Planck_lensing}.
\end{itemize}

For the sake of notation, we will refer to the combination of these 4 likelihoods as ``PL18''. Furthermore, recently the \textit{Planck} Collaboration has released \textit{Planck} Release 4 (PR4), exploiting the NPIPE end-to-end pipeline \cite{Mangilli:2015xya, Couchot_2017a, Couchot_2017b}. This instead consists of
\begin{itemize}
    \item ``High-L Likelihood Polarized for \textit{Planck}'' (HiLLiPoP)\footnote{\url{https://github.com/planck-npipe/hillipop}}, encoding the high-$\ell$ region of TT,TE and EE;
    \item ``Low-L Likelihood Polarized for \textit{Planck}'' (LoLLiPoP)\footnote{\url{https://github.com/planck-npipe/lollipop}}, encoding the low-$\ell$ one of EE, EB and BB.
\end{itemize}

Note that LoLLiPoP contains information on B-modes, whereas PL18 does not. With these additional likelihoods, we define another abbreviation, i.e. ``PL21''. This corresponds explicitly to \textit{plikTTTEEE}+\textit{lowlTT}+LoLLiPoP+\textit{lensing}, in such a way that the high multipole part is still carried out by the likelihood of PR3, together with the low-$\ell$ part of temperature and lensing. Instead, the low-$\ell$ part of E-mode, B-modes, and their cross-correlation is described through the PR4. In principle, PR4 is not independent of PR3, thus one has to be careful on combining them. In our case, we isolate the PR4 contribution to the low-$\ell$ part of the polarization fields (LoLLiPoP), so that we can still combine it with products of PR3 encoding different multipole ranges, or fields, assuming them to be independent.

\subsection{BICEP/Keck Array}

Together with \textit{Planck}, the BICEP/Keck array data have been crucial in obtaining our current knowledge on tensor perturbations. As for \textit{Planck}, the data are available through public likelihoods, named:

\begin{itemize}
    \item ``BK15'', representing the measurement of BICEP2/Keck Array of the B-mode polarization \citep{Ade_2018};
    \item ``BK18'', representing the newly released dataset from BICEP3/Keck Array  \citep{Tristram:2022}. 
\end{itemize}

\subsection{LIGO-Virgo-KAGRA interferometers}

As mentioned above, CMB alone can only probe the largest scales of the tensor primordial spectrum (approximately $k\simeq 10^{-2}$ Mpc$^{-1}$). In fact, even in the ideal case of no instrumental noise, the cosmological B-mode contribution would be several orders of magnitude lower than the lensing contribution on small scales (see figure \ref{fig: exp}). On the other hand, GW interferometers are probing scales almost 18 orders of magnitude away from the CMB ones (approximately $k\simeq 10^{16}$ Mpc$^{-1}$), thus they provide a way to strongly constrain small scales. This line of reasoning hides a caveat though: in order to obtain helpful information on the tilt from GW interferometers, we must assume that $n_t$ remains constant on the huge range of frequencies dividing CMB from interferometers, which could not be the case in nature \cite{Kuroyanagi_2011, Giare_2021a, Giare_2021b}. However, given that we have not yet detected the amplitude of the tensor modes, i.e. $r$, it is already challenging to constrain the tensor tilt $n_t$; thus, we choose to neglect any running of the tensor tilt, as commonly done in the literature. 

Currently, interferometers provide only upper bounds on the energy density of GWs at their frequency range. Indeed, knowing the minimal energy density detectable by a GW interferometer and assuming our parametrization of the tensor power spectrum (see eq. \ref{eq: power_spectra}), one can extract a bound on how blue tilted the primordial spectrum can be. In particular, one can translate $\qty(r_{0.01}, n_t)$ into an energy density at some reference frequency $f$ with \cite{Cabass_2016, Planck_2018}
\begin{equation}
    \Omega_{GW}(f)= \frac{r_{0.01} A_s}{24 z_{eq}}\qty(\frac{f}{f_{CMB}})^{n_t}\ ,
    \label{eq: omegagw}
\end{equation}
where $f_{CMB}$ is the frequency corresponding to the chosen pivot scale (e.g. $k=0.01$ Mpc$^{-1}$) and $z_{eq} \simeq 3400$ is the redshift of the matter-radiation equality \cite{Planck_2018}.

The datasets from GWs interferometers will be named:
\begin{itemize}
    \item ``LV15'', referring to the results after the first observing run of LVK. The upper bound on the energy density at a reference frequency of $f_{LVK} = 20$ Hz is $\Omega_{GW}(f_{LVK}) < 1.7 \times 10^{-7}$ at $95\%$ CL. Note that this was obtained by assuming a scale-invariant spectrum \cite{abbott_2017_LIGO}.
    \item ``LV18'', referring to the results after the second observing run of LVK. This time, the bound on the energy density we consider is marginalized on the value of the spectral tilt and at a frequency of 25 Hz. In particular, $\Omega_{GW}(25 \text{ Hz}) < 3.4 \times 10^{-8}$ at $95\%$ CL \cite{Abbott_2019_ligo}.
    \item ``LV21'', referring to the results after the third observing run of LVK. The upper bound reported is $\Omega_{GW}(25 \text{ Hz}) < 6.6 \times 10^{-9}$ at $95\%$ CL, obtained once again marginalizing over the spectral tilt \cite{ligo_meas}.
\end{itemize}

Notice that these datasets are not independent. Instead, they are the result of a longer observation time and of a progressive improvement of the systematics affecting LVK. For this reason, we cannot combine them since the information provided by LV15 is already contained in LV18, and so on.

As regards the actual inclusion of these bounds, one may encode them in an MCMC analysis as a half-Gaussian prior to the energy density of GWs predicted with eq. \ref{eq: omegagw}, having the $95\%$ limit at the value reported by LVK (this approach is used in \cite{Planck_2018}); an alternative, but equivalent way would be to define a Gaussian likelihood for LVK, centered in $\mu_{LVK} = 0$ and having a dispersion $\sigma_{LVK}$ that is half of the $95\%$ bound provided by LVK:
\begin{equation}
    -2\log\qty(\mathcal{L}_{LVK}) = \frac{\qty(\Omega_{GW} - \mu_{LVK})^2}{\sigma_{LVK}^2}\ ,
\end{equation}
where $\Omega_{GW}$ is the value extracted from the MCMC analysis for a specific set of parameters (this is how we include it in our analysis).

\subsection{NANOGrav}

Together with these data, we also want to study the consequences of the claim from NANOGrav collaboration \cite{nanograv_meas}. In fact, they report a significant detection of a common signal among several pulsars, but they do not obtain a clear detection of the spatial correlation of those, which would be the definitive proof of the stochastic origin of the signal. Despite this, it is interesting to include this dataset in our analysis, implying the assumption of a cosmological origin.

As regards the actual bound, there is an intermediate step we must take w.r.t. the LVK case. NANOGrav Collaboration reports its data in terms of amplitude $A_{CP}$ and spectral tilt $\alpha_{CP}$ (see \cite{nanograv_meas} for details); therefore, we must first obtain the corresponding $\Omega_{\rm GW}$. In particular \cite{MAGGIORE2000283, Guzzetti:2016mkm}:
\begin{equation}
    \Omega_{\rm GW}(f) = \frac{2\pi^2}{3H_0^2} f^2h_c^2(f)\ ,
\end{equation}
where $h_c$ is the power spectrum of the characteristic GW strain reading
\begin{equation}
    h_c(f) = A_{CP} \qty(\frac{f}{f_{yr}})^{\alpha_{CP}}\ .
\end{equation}

\begin{figure}[t]
    \centering
    \includegraphics[width=.5\hsize]{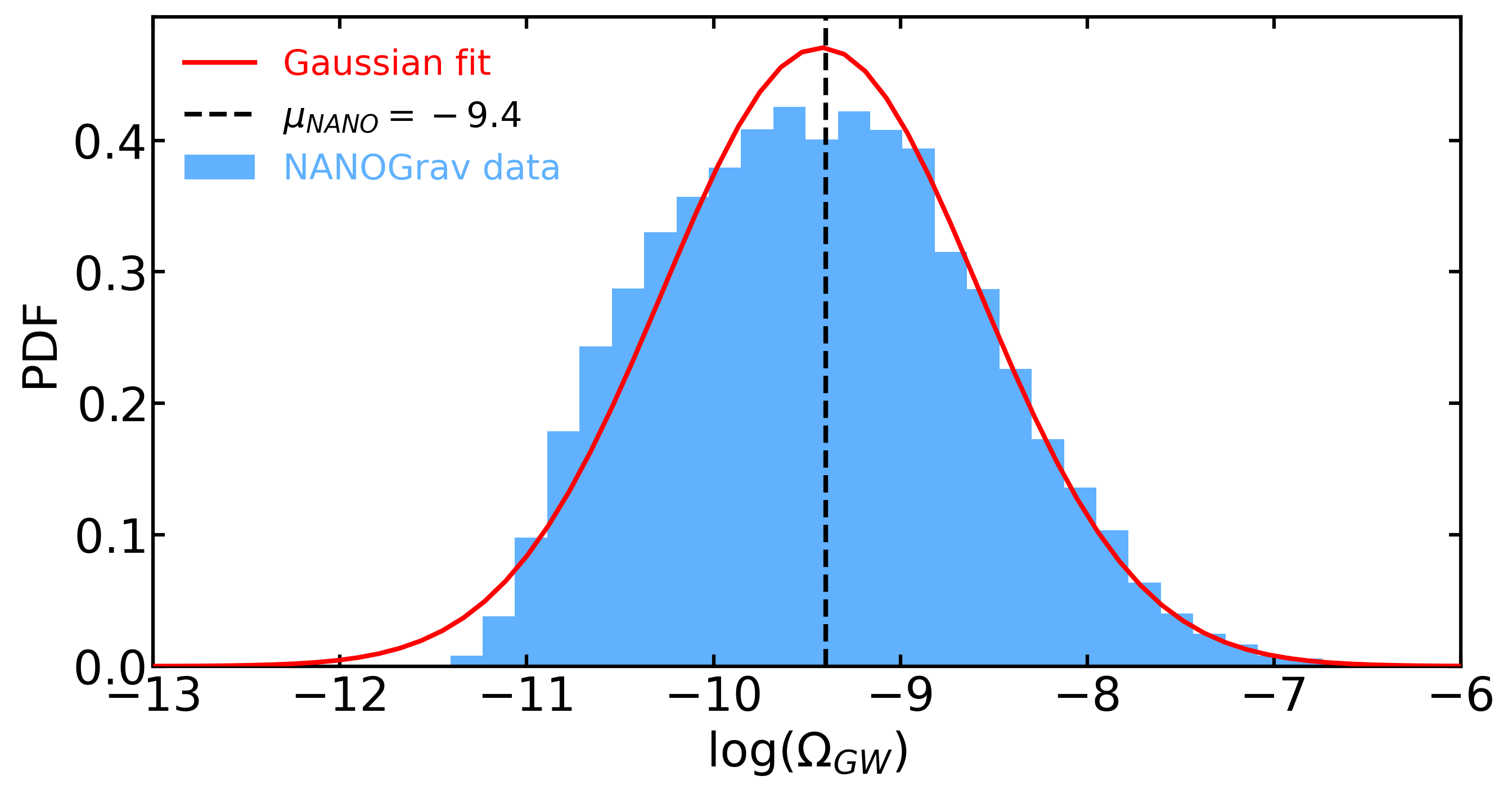}
    \caption{NANOGrav data marginalized over the spectral index. The solid curve represents the Gaussian likelihood we used for our analysis.}
    \label{fig: nano_data}
\end{figure}

Now, each $\qty{A_{CP}, \alpha_{CP}}$ couple will correspond to a function $\Omega_{\rm GW}(f)$. However, mimicking what is done for the LVK results, we are interested in the reference frequency of the experiment, thus we fix the frequency to $f = f_{NANO} = 1$ yr$^{-1}$ (which corresponds to $k_{NANO} \simeq 2.0 \times 10^{7}$ Mpc$^{-1}$), such that 
\begin{equation}
    \Omega_{\rm GW}(f_{yr}) = \frac{2\pi^2}{3H_0^2} f^2_{yr} A_{CP}^2\ .
\end{equation}
This does not depend on $\alpha_{CP}$; at this point, one can verify that the logarithm of $\Omega_{GW}$ follows approximately a Gaussian distribution, thus we fit it to find the bound: $\log\qty(\Omega_{\rm GW}) = -9.4 \pm 0.8$. This procedure allows us to get figure \ref{fig: nano_data}, which is obtained from the NANOGrav chains\footnote{The dataset and further information can be found at \url{https://github.com/nanograv/12p5yr_stochastic_analysis}.}.
To include this bound in the MCMC, one may follow the same procedure proposed for the LVK, thus defining a likelihood as
\begin{equation}
    -2\log \qty(\mathcal{L}_{NANO}) = \frac{\qty[\log\qty(\Omega_{\rm GW}) - \mu_{NANO}]^2}{\sigma_{NANO}^2}\ .
\end{equation}
Differently from the LVK case, now $\mu_{NANO} \neq 0$. Finally, to refer to this dataset we will use ``NANO''.

We collect all the names and abbreviations in table \ref{tab: names}. 

\begin{table*}
\centering
\addtolength{\leftskip} {-3.3cm}
\addtolength{\rightskip}{-3cm}
\begin{tabular}{l@{\hskip 0.05in}c@{\hskip 0.05in}l}
    \toprule
    \textbf{Name} & \textbf{Likelihoods}  & \textbf{Description} \\ \midrule
    PL18          & plikTTTEEE+lowlTT+lowlEE+lens. & Combination from PR3 \cite{Planck_like, Planck_lensing} \\
    PL21          & plikTTTEEE+lowlTT+LoLLiPoP+lens.     & Combination of PR3 and PR4 \cite{Planck_like, Planck_lensing, Couchot_2017a} \\
    BK15          & BICEP2/Keck array        & BICEP2/Keck Array likelihood \cite{Ade_2018}   \\
    BK18          & BICEP3/Keck array       & BICEP3/Keck Array likelihood \cite{BICEP_2021}   \\
    LV15          & LIGO-Virgo-KAGRA       & LVK likelihood with 2015 data \cite{abbott_2017_LIGO} \\
    LV18          & LIGO-Virgo-KAGRA       & LVK likelihood with 2018 data \cite{Abbott_2019_ligo} \\
    LV21          & LIGO-Virgo-KAGRA       & LVK likelihood with 2021 data \cite{ligo_meas} \\
    NANO          & NANOGrav               & NANOGrav likelihood with 2021 data \cite{nanograv_meas} \\ \bottomrule
\end{tabular}
\caption{Names of the main likelihoods, or collection of them, used in this work.}
\label{tab: names}
\end{table*} 

\section{State-of-the-art}\label{sec: state}

In this section, we will briefly go through the combination of datasets that have been implemented to obtain the current bounds on $r_{0.01}$ (or $r_{0.05}$) and $n_t$ (the literature on this is quite vast, thus we consider only a few of the most recent works). 

\citet{Planck_2018} (\textit{Planck} 2018) consider two combinations while studying the $\Lambda$CDM+$r_1$+$r_2$ model:
\begin{itemize}
    \item PL18 + BK15, obtaining $r_{0.01}<0.076$ and $-0.55 < n_t < 2.54$ at $95\%$ CL, obtained by exploiting TSA with $(k_1,k_2) = (0.002,0.02)$ Mpc$^{-1}$. 
    \item PL18 + BK15 + LV15, resulting in $r_{0.01} < 0.066$ and $-0.76 < n_t < 0.52$ at $95\%$ CL, using TSA with $(k_1,k_2) = (0.002,0.02)$ Mpc$^{-1}$. 
\end{itemize}

From this, it is clear that CMB experiments are not able to constrain very well the bluest tilts of the spectrum. Instead, the addition of LV15 (as a half-Gaussian prior) severely cuts the allowed range of $n_t$. Notice that LV15 is obtained by fixing the spectral tilt to scale invariance, while the MCMC is performed by letting it vary.

\citet{Tristram2021} explored the PR4 on the $\Lambda$CDM+$r_{0.05}$ model, thus keeping $n_t$ fixed to its single-field slow-roll prediction: $n_t = -r_{0.05}/8$, where $0.05$ Mpc$^{-1}$ is the pivot scale of scalar perturbations. Given that they report their results for this scale, we will compute a very rough estimate of the corresponding bound on $r_{0.01}$. In particular, if they report $r_{0.05} < X$ at $95\%$ CL, we will use eq. \ref{eq: coordinate}, $n_t = -X/8$, and the best-fit value of $n_s$ of \textit{Planck 2018} to get the corresponding $r_{0.01}<X^\prime$ bound (we will report this in parenthesis). The most relevant (for this work) cases analyzed in \cite{Tristram2021} are:
\begin{itemize}
    \item LoLLiPoP, which yields $r_{0.05} < 0.069$ at $95\%$ CL ($r_{0.01} < 0.066$).
    \item HiLLiPoP(only TT)+\textit{lowlTT}+LoLLiPoP, yielding $r_{0.05} < 0.060$ at $95\%$ CL ($r_{0.01} < 0.057$).
    \item HiLLiPoP(only TT)+\textit{lowlTT}+LoLLiPoP+BK15, yielding $r_{0.05} < 0.044$ at $95\%$ CL ($r_{0.01} < 0.042$).
\end{itemize}

\citet{BICEP_2021} presented the new release of data from the BICEP/Keck Array (here BK18). While keeping the $\Lambda$CDM parameters fixed and using $n_t = -r_{0.05}/8$, they obtain $r_{0.05}<0.036$ at $95\%$ CL ($r_{0.01} < 0.034$).

Finally, \citet{Tristram:2022} used PR4 and BK18 to obtain the state-of-the-art upper bound on $r_{0.05}$ on the $\Lambda$CDM+$r_{0.05}$ model ($n_t = -r_{0.05}/8$). In particular, they consider HiLLiPoP + \textit{lowlTT} + LoLLiPoP + BK18 + BAO + \textit{lensing}, finding $r_{0.05} < 0.032$ at $95\%$ CL ($r_{0.01} < 0.030$).
Notice that, they consider a dataset we are not including in this work, which is Baryonic Acoustic Oscillations (BAO) from \cite{Alam_2021}.

This set of bounds, especially \citet{Tristram:2022} for what regards $r_{0.05}$, and \citet{Planck_2018} for what regards $n_t$, represent the current state-of-the-art. As mentioned before, for the ones with $n_t$ fixed, we compute in parenthesis a corresponding $r_{0.01}$ value, however, comparing them with the results we are about to show is not so trivial. In fact, we let $n_t$ vary, thus our results on $r_{0.01}$ will be marginalized on the tilt. However, that rough correspondence can help put this work in perspective w.r.t the rest of the literature.

\section{Results}\label{sec: Results}

Finally, let us report the results of our analysis. This has been performed using the SSA on the complete $\Lambda$CDM+$r_{0.01}$+$n_t$ parameter space, assuming $r_{0.01}^{cut} = 10^{-5}$. The priors for these parameters are flat and their ranges are shown in table \ref{tab: priors} (in appendix \ref{sec: TSA} we also show the results for the TSA case). As for the convergence, we used the Gelman-Rubin test to determine when to stop the MCMC runs \cite{Gelman_Rubin}. Given the number of dataset combinations, the number of varied parameters making the convergence difficult to achieve, and the finite CPU hours available, we assumed a minimal level of convergence of $R-1<0.04$. For a few selected cases (the most constraining ones), we instead assumed $R-1<0.01$ in order to explore more thoroughly the tails of the distribution and to avoid over/underestimating our bounds. As regards the size of the MCMC chains, we quantify it using the ``total weights'' given by \texttt{GetDist}. Indeed, when a proposed step of the chain is rejected, the weight of the current point in parameter space will increase, as if it is counted once again. In other words, this quantity represents the total number of steps, regardless of their being accepted, or not. As an example, the shortest length obtained (PL18+BK18) is $\sim27,000$, while the longest is $\sim141,000$ (PL21+BK18+LV21).

We performed the MCMC runs for most of the relevant permutations of \textit{Planck}, BICEP/Keck, and LVK data, searching for the most constraining one. Instead, given the nature of the NANOGrav claim, and knowing the results of our analysis, we keep the discussion on that dataset separated from the others. 

Table \ref{tab: 2D_res} reports the one-dimensional $95\%$ CL for each combination analyzed. Assuming PL18 + BK15 to be our ``starting point''\footnote{As mentioned before this is essentially the state-of-the-art combination for what regards the case in which $n_t$ is allowed to vary. Notice that here we are not including LV15, since it is actually obtained assuming scale-invariance at the typical frequency range of interferometers.}, the first 4 rows show the individual improvements brought by various datasets. Among them, it is clear that BK18 prominently improves the bound on the tensor-to-scalar ratio (as shown by \cite{BICEP_2021, Tristram:2022}), while shifting the allowed range of $n_t$ towards lower values w.r.t. PL18+BK15.

The next 4 cases feature combinations of different upgrades w.r.t. PL18+BK15. Given that we want to privilege the stricter bound on the tensor-to-scalar ratio, PL21$+$BK18$+$LV21 is what we consider our best dataset to constrain the tensor sector, resulting in 
\begin{equation}
        r_{0.01} < 0.028  \qq{and} -1.37 < n_t < 0.42 \ ,
\end{equation}
at $95\%$ CL, which is the main result of this work and represents the most constraining bound on the tensor-to-scalar perturbation ratio to our knowledge. 

\begin{figure}[t]
    \centering
    \includegraphics[width=.5\hsize]{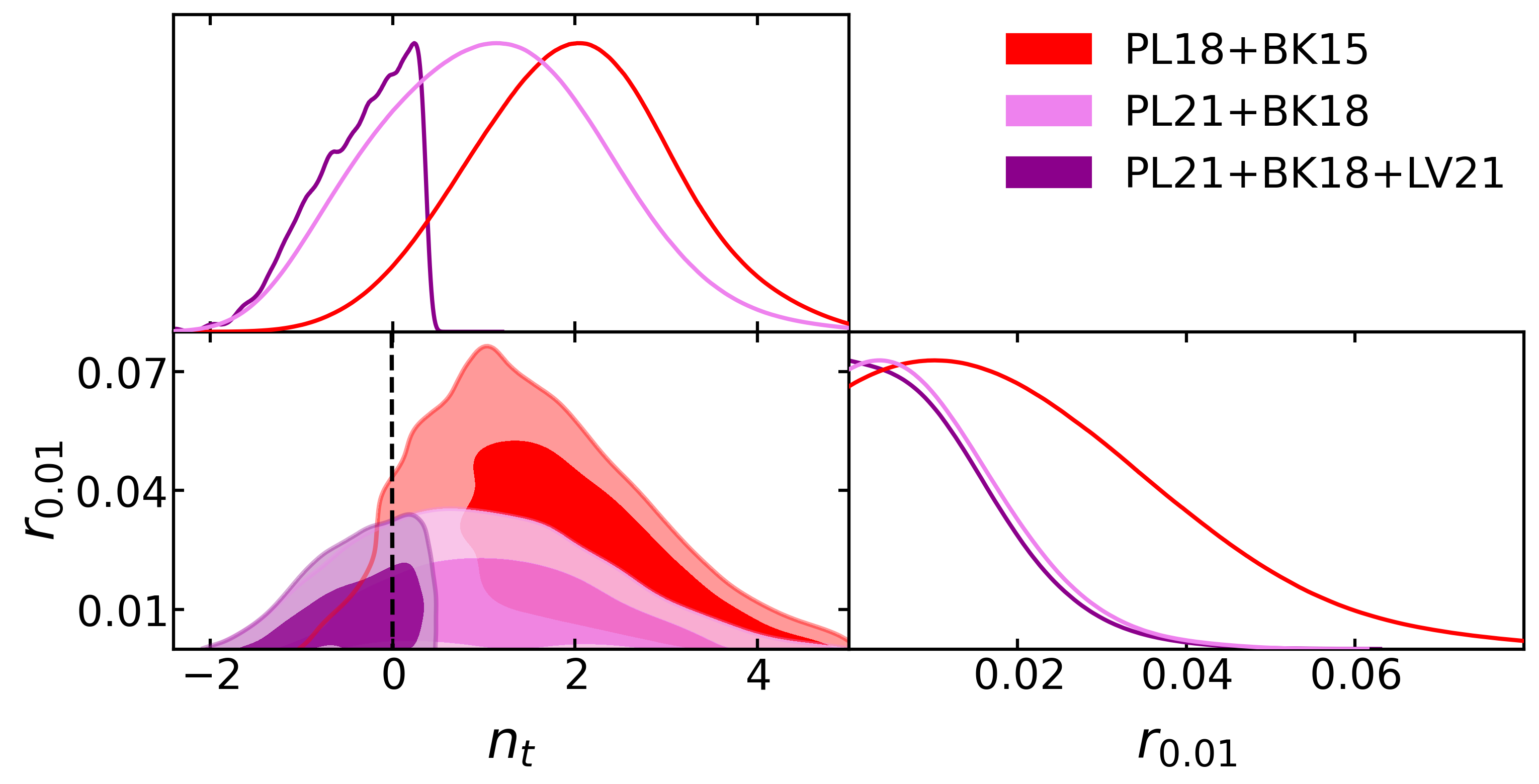}
    \caption{2D $68$ and $95\%$ CL contours in the $\qty(r_{0.01}, n_t)$-plane for PL18$+$BK15, PL21$+$BK18 and PL21$+$BK18$+$LV21. The dashed black line is the well-known slow-roll single-field prediction $n_t = -r/8 = -2 \epsilon$.}
    \label{fig: 2D_res}
\end{figure}

\begin{figure}[t]
    \centering
    \includegraphics[width=.7\hsize]{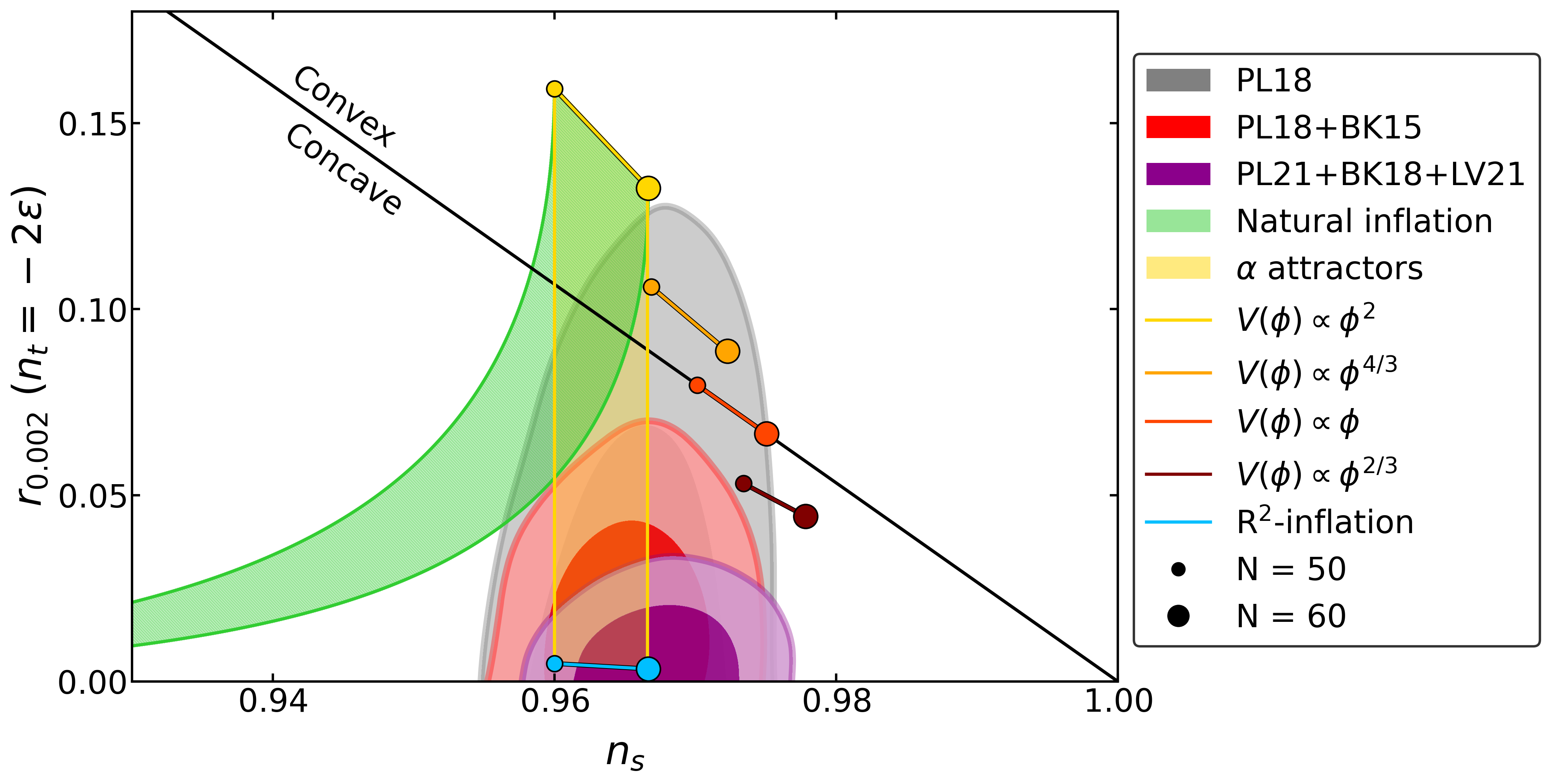}
    \caption{2D $68$ and $95\%$ CL intervals in the $\qty(r_{0.002}, n_s)$-plane for PL18 (publicly available MCMC chains\protect\footnotemark), PL18$+$BK15 and PL21$+$BK18$+$LV21. $r_{0.002}$ is obtained from our chains assuming the standard prediction $n_t = -r/8 = -2 \epsilon$. For more details on the various inflationary models, see \cite{Planck_2018}.}
    \label{fig: rns}
\end{figure}

We show the 2D posterior contours of this case in figure \ref{fig: 2D_res}, compared to some other relevant cases: PL18 + BK15 (representing the state of the art) and PL21 + BK18 (to underline the improvement carried out by LV21).
\begin{table}
  \footnotesize
  \centering
  \begin{tabular}{l c , .}
   \toprule
    & \multicolumn{1}{c}{$\mathbf{r_{0.01}}$ \textbf{95\% CL}}  & \multicolumn{1}{c}{$\mathbf{n_t}$ \textbf{95\% CL}} & \multicolumn{1}{c}{$\mathbf{R-1}$ \textbf{test}} \\ \midrule
   PL18+BK15       & < 0.056 & [-0.22,\ 4.16] & 0.032 \\
   PL18+BK18       & < 0.032 & [-0.98,\ 3.46] & 0.033 \\
   PL18+BK15+LV18  & < 0.059 & [-1.00,\ 0.45] & 0.039 \\
   PL18+BK15+LV21  & < 0.057 & [-0.91,\ 0.42] & 0.025 \\
   PL18+BK18+LV21  & < 0.032 & [-1.14,\ 0.42] & 0.034 \\
   PL21+BK15       & < 0.049 & [-0.60,\ 4.34] & 0.010 \\
   PL21+BK18       & < 0.029 & [-1.21,\ 3.54] &  0.016\\
   PL21+BK18+LV21  & < 0.028 & [-1.37,\ 0.42] & 0.006 \\ \hline
   PL18+BK15+NANO  & < 0.071 & [0.44,\ 0.83] & 0.028 \\
   PL21+BK18+NANO  & < 0.033 & [0.47,\ 0.85] & 0.005 \\ \bottomrule
  \end{tabular}
  \caption{95\% CL intervals of the 10 considered combinations of datasets. Our main result is PL21+BK18+LV21. Here we also show the results of the Gelman-Rubin test for each combination.}
  \label{tab: 2D_res}
\end{table}
Furthermore, in figure \ref{fig: rns} we show the posterior distribution on the plane $r_{0.002}-n_s$ for the same datasets as in figure \ref{fig: 2D_res}. To do so, we mimic figure 8 of \cite{Planck_2018}, showing a few relevant inflationary scenarios and fixing $n_t$ to its standard single-field prediction. As already highlighted by \cite{Tristram:2022, BICEP_2021}, we confirm with improved sensitivity that all potentials for the single-field inflationary models we show are ruled out by the data. The same can be said of natural inflation, which is well outside the 95\% CL region.
\setcounter{footnote}{2}
\footnotetext{\url{http://pla.esac.esa.int/pla/\#home}}

\subsection{Marginalization effects}
In section \ref{sec: MCMC} we show that marginalization effects should be taken into account to check the sanity of our results. Instead of performing a full PL analysis, here we show the Profile Distribution (PD) for our most constraining dataset, i.e. PL21+BK18+LV21. In fact, \citet{gomez-valent2022FastTestAssessimpact} presented this technique as an approximation of the PL useful if one already has an MCMC to exploit and proved its robustness on highly dimensional likelihoods, such as \textit{Planck}'s.
To obtain PDs on $r_{0.01}$ and $n_t$, we follow the following procedure: 
\begin{enumerate}
    \item starting from our MCMC chains, we bin the values of $r_{0.01}$ or $n_t$;
    \item in each bin, we search for the minimum value of $\chi^2$.
    \item Then, we fit a polynomial\footnote{In the case of a Gaussian likelihood this would simply be a parabola, which is the case for $r_{0.01}$. For the $n_t$ distribution, which is highly non-Gaussian, we use a polynomial with a degree equal to 8.}, using the mean value of the profiled parameter in the bin as the x-coordinate;
    \item the resulting curve represents $\chi^2$ as a function of this parameter.
\end{enumerate}
   
This procedure indeed approximates an actual PL, which instead is performed point-by-point via some minimization algorithm over all the non-fixed parameters. In this case, instead, the very low computation time of this technique (of the order of a fraction of a second) comes at the cost of having to choose a binning strategy. In fact, the value of $\chi^2$ in each bin is ``noisy'', since it is not guaranteed that it has reached its absolute minimum. Clearly, the larger the bin, the more probable it is to find the real minimum of $\chi^2$, and the larger the uncertainty on the x-axis. In other words, one has to find the right trade-off between having small bins and populating them sufficiently. In our case, we focus mainly on $r_{0.01}$, for which we explore 4 different binning strategies in the range $[0, 0.05]$ with 10, 30, 50, and 100 bins. Instead, we bin $n_t$ in $[-2.0,0.48]$ with 100 bins.
\begin{figure}[t]
    \centering
    \includegraphics[width = .49\textwidth]{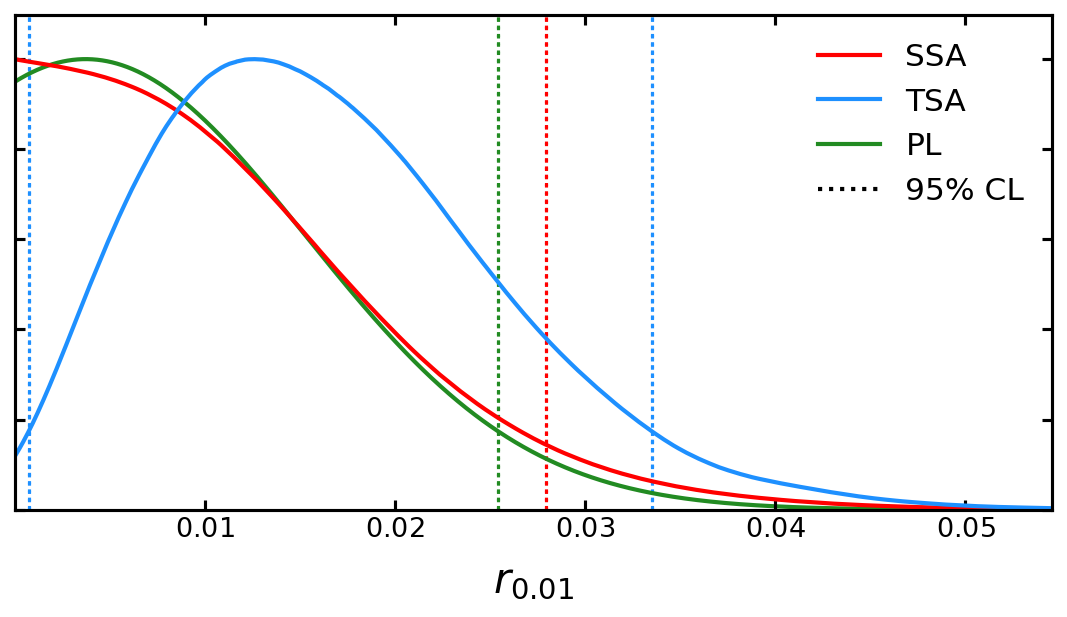}
    \includegraphics[width = .49\textwidth]{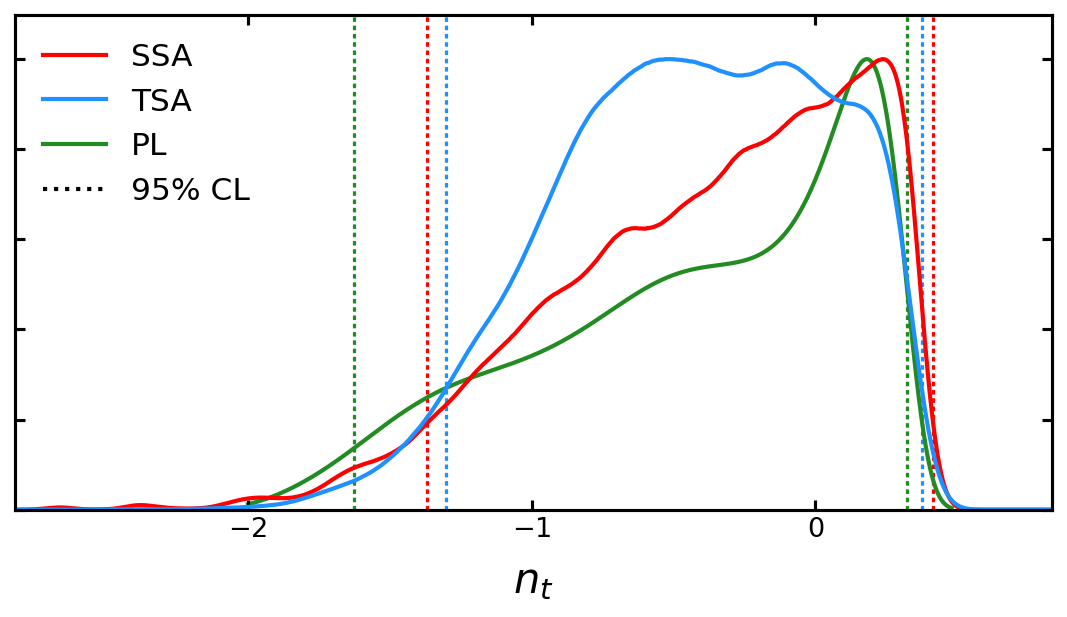}
    \caption{1D results on $r_{0.01}$ and $n_t$, obtained using SSA, TSA, and PL on our most constraining dataset, that is, PL21+BK18+LV21. This plot refers to the case with 30 bins. The vertical dotted lines represent the boundaries of the SSA and TSA 95\% CL intervals. Note that SSA gives a single-tail bound on $r_{0.01}$ while the TSA provides a two-tail one. Also, the green dotted lines indicate the 95\% credible intervals of the PD. These are also summarized in table \ref{tab: data_MCMC_vs_PL}.}
    \label{fig: data_PLvsMCMC}
\end{figure}
\begin{table}[t]
\centering
\begin{tabular}{ccc}
    \toprule
     & $\mathbf{r_{0.01}}$ \textbf{95\% CL} & $\mathbf{n_t}$ \textbf{95\% CL} \\ 
    \midrule
    SSA & $< 0.028$ & ${[}-1.37,\ 0.42{]}$ \\
    TSA & ${[}0.01,\ 0.033{]}$ & ${[}-1.30,\ 0.37{]}$ \\
    \midrule
    & $\mathbf{r_{0.01}}$ \textbf{95\% credible interval} & $\mathbf{n_t}$ \textbf{95\% credible interval} \\ 
    \midrule
    PD  & $< 0.025$ & ${[}-1.63,\ 0.32{]}$ \\ 
    \bottomrule
\end{tabular}
\caption{$95\%$ CL obtained with SSA and the TSA compared to the 95\% credible intervals of the PD.}
\label{tab: data_MCMC_vs_PL}
\end{table}

Figure \ref{fig: data_PLvsMCMC} shows the results for $r_{0.01}$ (30 bins) and $n_t$, respectively (see also table \ref{tab: data_MCMC_vs_PL}). In terms of 95\% credible intervals, we obtain $-1.63<n_t<0.32$. For $r_{0.01}$ we obtain $r_{0.01}< 0.028, 0.025, 0.025$, and $0.024$ using 10, 30, 50, and 100 bins, respectively. Also, we compare the PL with both the SSA and the TSA results. The left panel shows that the SSA and the PL are very similar, even if the SSA seems to obtain slightly more conservative results. This shows that marginalization effects have only a partial role in obtaining our result for PL21+BK18+LV21. Note that the case with 10 bins, thus the one in which it is more probable that $\chi^2$ reaches its absolute minimum, gives the same result as our MCMC. 

On the other hand, it is clear that the bias observed in the TSA case is due to the informative prior discussed in section \ref{sec: test_prior}. 

Looking at the right panel, both the SSA and TSA are fairly similar to the PL distribution, with the former being slightly more conservative, as expected.

\subsection{Including NANOGrav}

In order to study the posterior distribution of $r_{0.01}$ and $n_t$ when including the NANOGrav dataset, we do not make use of either LV18 or LV21. Also, let us recall that performing this analysis for NANOGrav implies the assumption of a cosmological origin of the measured signal.  

The last two rows of Tab.\ref{tab: 2D_res} (see also figure \ref{fig: bounds}) show the results of PL18$+$BK15$+$NANO and PL21$+$BK18$+$NANO, together with all other 8 combinations. Notice that there is a discrepancy between the estimated 1D 95\% CL bound of $n_t$ obtained exploiting NANOGrav or LVK. The reference frequency of NANOGrav is $f_{NANO} = 1 $ yr$^{-1} \simeq 32\times$ nHz $\to k_{NANO} \simeq 2.0 \times 10^{7}$ Mpc$^{-1}$, instead of the LVK one is $f_{LVK} = 25$ Hz $\to k_{LVK} \simeq 1.6 \times 10^{16}$ Mpc$^{-1}$, thus they are separated by 9 orders of magnitude. This discrepancy suggests that the signal detected by the NANOGrav collaboration cannot be a cosmological background signal characterized by a simple power-law \cite{Kawasaki_2021, Fujita_2022, Benetti_2022}. In fact, fitting the NANOGrav bound on $\Omega_{GW}$ at $k_{NANO}$ with our best-fit power-law of PL21$+$BK18$+$NANO would make it inconsistent with the LVK bound we have at $k_{LVK}$.

This is consistent with what has already been found in the literature \cite{Kawasaki_2021, Fujita_2022, Benetti_2022}. In fact, there is a zoology of different models trying to explain the NANOGrav claim in a cosmological fashion; however, this is beyond the scope of this work.

\section{Conclusions} \label{sec: conclusions}

In this work, we have obtained new bounds on the tensor-to-scalar ratio $r$ and the tensor spectral index $n_t$. We have exploited newly released datasets from both an electromagnetic point of view, i.e. BICEP/Keck 2018 \cite{BICEP_2021} and \textit{Planck} PR4 \cite{Tristram2021}, and a GW perspective, that is, LVK collaboration data \cite{ligo_meas}. In particular, the complementarity of \textit{Planck} and BK measurements allows us to better constrain the amplitude of the tensor perturbation spectrum, while the information at small scales coming from LVK can cut the values of permitted spectral tilts.

To obtain reliable bounds on the tensor sector, we have studied the behavior and performances of two approaches encoding $r$ and $n_t$ into our MCMC analysis, i.e. the SSA and the TSA. The former consists of sampling directly these two parameters while cutting the lowest values of $r$ at some undetectable threshold $r_{0.01}^{cut}$ \cite{Cabass_2016}; in the latter, one defines two tensor-to-scalar ratios $r_1$ and $r_2$ (at scales $k_1$ and $k_2$) with which the parameter space is explored \cite{Planck_2018}. Then $r_{0.01}$ and $n_t$ are recovered as functions of $r_1$ and $r_2$ (see eq. \ref{eq: coordinate}). In section \ref{sec: MCMC} we perform different tests to assess how a priori and arbitrary choices affect both methods. In particular, we inspect the prior information injected in the TSA by the coordinate transformation (see also appendix \ref{sec: coords}) and the performance of both approaches on a mock dataset. Furthermore, we make use of the PL technique to inquire about the presence of marginalization effects. 

Starting from the TSA, in section \ref{sec: MCMC} we find that the informative prior on $r_{0.01}$ biases the results toward higher values w.r.t. what data suggest (thus the PL results) to the point that $r_{0.01}=0$ is excluded at more than 95\% CL. In addition, the informative prior on $n_t$ introduces a strong pull toward scale-invariance ($n_t=0$) disfavoring inflationary models characterized by a large $\abs{n_t}$. 

With respect to SSA, we find that the main weakness is marginalization effects. Thus, even if one considers a very low $r_{0.01}^{cut}$ and a large prior in $n_t$ (thus without excluding a priori a large part of the posterior volume), there might be a push towards small values of $r_{0.01}$, caused by the presence of a large posterior volume near $r_{0.01}$. 

Despite this, this effect can be gauged using the PL, yielding that the SSA seems to be more consistent with our PL results w.r.t. to the TSA (see section \ref{sec: test_exact}). Thus, we provide our baseline results in section \ref{sec: Results} using the SSA (appendix \ref{sec: TSA} reports the TSA results).

We have analyzed 10 combinations of the available datasets and we have found that the most constraining consists of a combination of \textit{Planck} PR3 (high-$\ell$ part of TT, EE, and TE, low-$\ell$ part of TT and lensing) and PR4 (low-$\ell$ part of EE, BB, and EB), BK18, and the last release from LVK. Together they provide $r<0.028$ and $-1.37 < n_t < 0.42$ at 95\% CL with a sensitivity on $r$ of $\sigma_r = 0.0086$. To our knowledge, these are the most constraining bounds available in the literature for what concerns an inflationary scenario characterized by a power-law. Standard single-field slow-roll prediction for the spectral tilt ($n_t = -r/8 = -2\epsilon$) is still completely compatible with our results. The TSA gives $0.001 < r < 0.033$ and $-1.32 < n_t < 0.37$ at 95\% CL on the same combination of datasets, which seems to be consistent with the biases mentioned before.

Furthermore, to assess the possible influence of marginalization effects on our main result with the SSA, in section \ref{sec: Results} we compute the PD of our MCMC chain using the technique proposed in \cite{gomez-valent2022FastTestAssessimpact}, which approximates the results of a PL. The agreement between this and our SSA marginalized distributions suggests that our results are driven by data and not by marginalization effects, which surely play a role, but not a significant one. In fact, for each binning strategy, we assumed, the upper bound of $r_{0.01}$ provided by the PD is always less than or equal to $r_{0.01} < 0.028$, demonstrating the robustness of our result. Instead, the biases on the TSA find confirmation when comparing it with the PD.

In addition, we have considered two additional combinations of data sets that account for the NANOGrav collaboration results \cite{nanograv_meas}. In these cases, we found an apparent inconsistency in the allowed range of $n_t$ w.r.t. what we have obtained using LVK data as our small-scale dataset, according to what has already been underlined in the literature. Specifically, we have obtained $r<0.033$ and $0.47 < n_t < 0.85$ at 95\% CL. If we assume that NANOGrav common-signal has a cosmological origin, one must abandon the single power-law description of the tensor power spectrum, given that NANOGrav and LVK cannot be reconciled in such a context. Instead, one may consider Axion-SU(2) spectator field \cite{campeti_2020_measuring, Kawasaki_2021, Fujita_2022}, broken power-law description of the primordial tensor spectrum \cite{Benetti_2022}, GW contribution to the relativistic degrees of freedom \cite{Vagnozzi_2020, Kuroyanagi_2021}, k-dependence of the tensor tilt \cite{Kuroyanagi_2011, Giare_2021a, Giare_2021b}, cosmic strings \cite{Ellis_2021}, phase transitions \cite{Ashoorioon2022}, domain walls or large-amplitude curvature perturbations \cite{Bian_2021} to obtain a framework accounting for all constraints at all scales.

In the future, CMB space missions such as LiteBIRD \cite{Hazumi:2019lys} will provide an unprecedented capability of measuring the large-scale part of the B-mode spectrum. Indeed, the target sensitivity of LiteBIRD alone is $\sigma_r = 0.001$ (a factor $\sim 9$ better than the sensitivity obtained in this work); thus it will allow us to obtain crucial information of the inflationary model which took place in nature (for some insights, see \cite{PTEP_LiteBIRD}).

\begin{appendix} 

\section{Constraints on the \texorpdfstring{$\Lambda$}{TEXT}CDM parameters} \label{sec: LCDM}
\begin{figure*}
    \centering
    \includegraphics[width=\hsize]{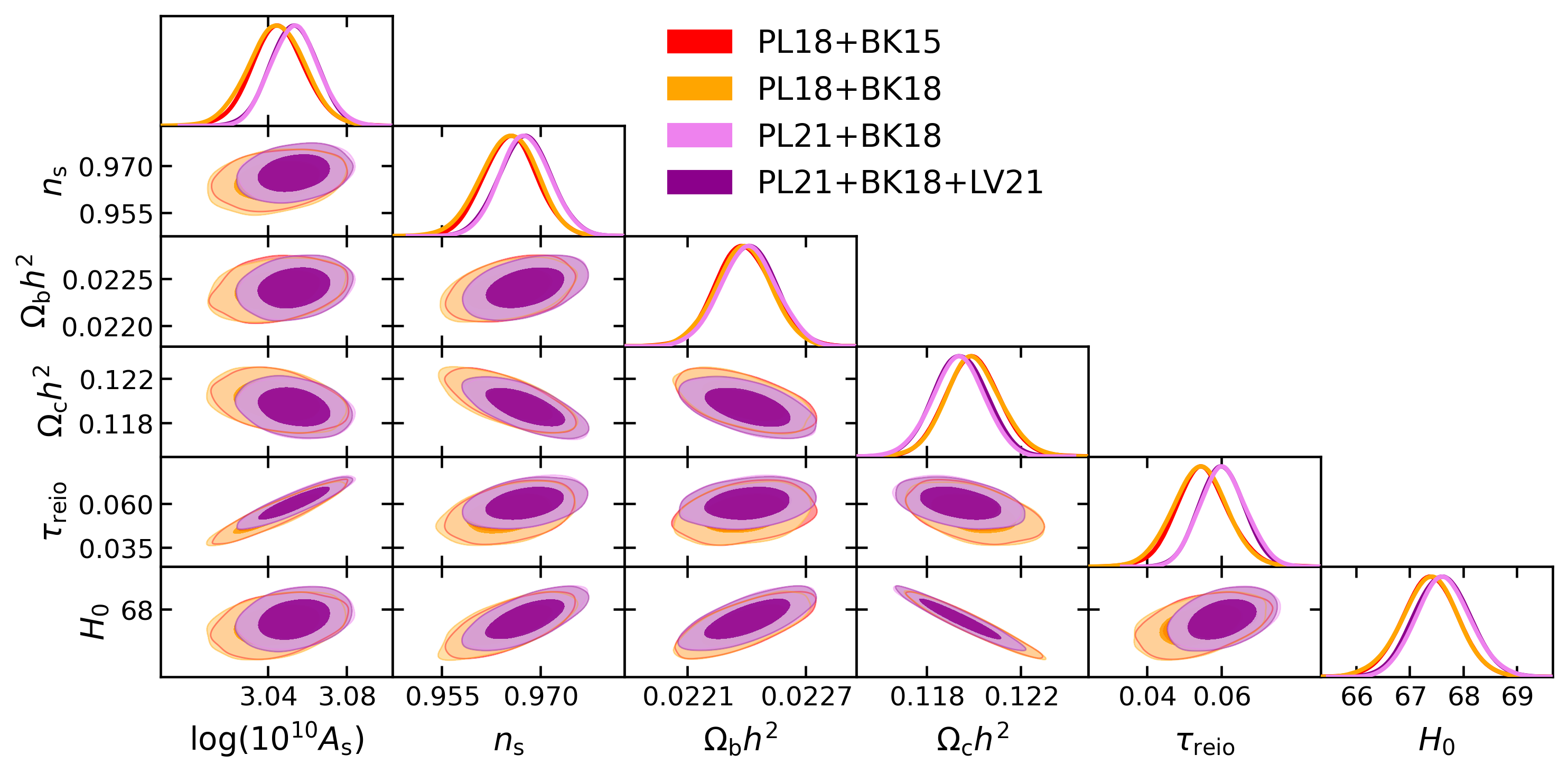}
    \caption{Posterior distribution for $\qty{\log(10^{10}A_s), n_s, \Omega_bh^2, \Omega_{cdm}h^2, \tau_{reio}, H_0}$.}
    \label{fig: LCDM}
\end{figure*}

Figure \ref{fig: LCDM} (see also Tab.\ref{tab: LCDM} for mean values and standard deviations) shows the posterior distribution for the 6 $\Lambda$CDM parameters for the most constraining combination for the tensor sector, i.e. PL21$+$BK18$+$LV21, compared to PL18$+$BK15, PL18$+$BK18, and PL21$+$BK18. The results shown here refer to our baseline approach to the tensor sector, i.e. SSA (see the beginning of section \ref{sec: Results} for a summary of the assumptions used).
\begin{table*}
\centering
  \begin{tabular}{l c c c c}
   \toprule
   \textbf{\textbf{Parameter}} & \textbf{PL18$+$BK15} & \textbf{PL18$+$BK18} & \textbf{PL21$+$BK18} & \textbf{PL21$+$BK18$+$LV21} \\  \midrule
    $\log\qty(10^{10}A_s)$ & $3.046 \pm 0.014$     & $3.044 \pm 0.014$     & $3.053 \pm 0.012$     & $3.053 \pm 0.012$     \\
    $n_s$                  & $0.9654 \pm 0.0040$   & $0.9652 \pm 0.0042$   & $0.9676 \pm 0.0039$   & $0.9676 \pm 0.0039$  \\
    $\Omega_bh^2$          & $0.02239 \pm 0.00015$ & $0.02238 \pm 0.00014$ & $0.02241 \pm 0.00014$ & $0.02241 \pm 0.00014$ \\
    $\Omega_{cdm}h^2$      & $0.1200 \pm 0.0012$   & $0.1200 \pm 0.0012$   & $0.1194 \pm 0.0011$   & $0.1194 \pm 0.0011$   \\
    $\tau_{reio}$          & $0.0549 \pm 0.0073$   & $0.0542 \pm 0.0074$   & $0.0603 \pm 0.0061$   & $0.0602 \pm 0.0061$  \\
    $H_0$                  & $67.38 \pm 0.53$      & $67.37 \pm 0.53$      & $67.62 \pm 0.51$      & $67.61 \pm 0.52$ \\ \bottomrule
  \end{tabular}
\caption{Mean and standard deviation of the 6 $\Lambda$CDM parameters using a selection of the analyzed dataset.}
\label{tab: LCDM}
\end{table*}
Looking at the first two columns, we know that the constraints on the 6 $\Lambda$CDM parameter are essentially driven by PL18. Thus, we can compare our results with \textit{Planck} 2018 ones \cite{Planck_parameters}, which are extremely compatible with ours (almost identical). 

Instead, it is less trivial to compare the other two columns with the literature, given that, to our knowledge, our combination of datasets PL21 has not been explored. Despite this, we know that the main differences are caused by the fact that LoLLiPoP provides a better characterization of the low-$\ell$ part of the polarization, while HiLLiPoP gives almost identical results w.r.t. PR3 \cite{Tristram2021}. Then, we can, for example, compare table \ref{tab: LCDM} with \cite{Campeti:2022vom}. They report the best-fit values on the $\Lambda$CDM parameters (together with all the others) obtained using HiLLiPoP + \textit{lowlTT} + LoLLiPoP + BK18 + BAO (\cite{Alam_2021}) + \textit{lensing}. These values are almost identical to the mean values we obtain from the MCMC; thus, the two are extremely compatible given our error bars. Since the difference w.r.t. \textit{Planck} 2018 reported in \cite{Campeti:2022vom} is due to LoLLiPoP, we can confirm the same shifts; however, note that the results of \cite{Campeti:2022vom} are obtained through a PL. 

The general behavior of our results can also be found in \cite{Tristram2021}. In particular, it seems that the major effect of PR4 is to push for slightly higher values of $\qty{\log\qty(10^{10}A_s),n_s,\tau_{reio}}$ w.r.t. \textit{Planck} 2018. However, notice that they do not exploit the high-$\ell$ part of polarization and lensing; therefore, one has to be careful in comparing \cite{Tristram2021} with this work.  

In conclusion, \citet{Planck:2016mks} constrains the reionization history using \textit{plikTT} + \textit{lowlTT} + LoLLiPoP + \textit{lensing}, finding $\tau_{reio} = 0.058^{+0.011}_{-0.012}$. This combination is very similar to the one used here (PL21+BK18), the difference is the high-$\ell$ part of polarization and BK18, and the result is compatible with ours. This confirms the shift we find in this parameter, which probably is also driving the increase in $\log\qty(10^{10}A_s)$ through their degeneracy. Also, $n_s$ and $H_0$ are positively correlated with $\tau_{reio}$ (although not as strongly as $\log\qty(10^{10}A_s)$), thus also their slight shift may be explained by the pull on the optical depth (at least partially). The opposite seems to happen to $\Omega_c h^2$, which is anti-correlated with $\tau_{reio}$.

\section{Coordinate transformation}\label{sec: coords}

In section \ref{sec: MCMC}, we perform a test to assess the prior information injected by the coordinate transformation $\qty(r_1-r_2) \to \qty(r_{0.01}-n_t)$ characterizing the TSA. To do so, we performed an MCMC analysis on priors alone, finding that the TSA has a strong pull toward scale invariance ($n_t=0$) and tends to exclude $r_{0.01}=0$. 

\begin{figure}[tbp]
    \centering
    \includegraphics[width = .45\hsize]{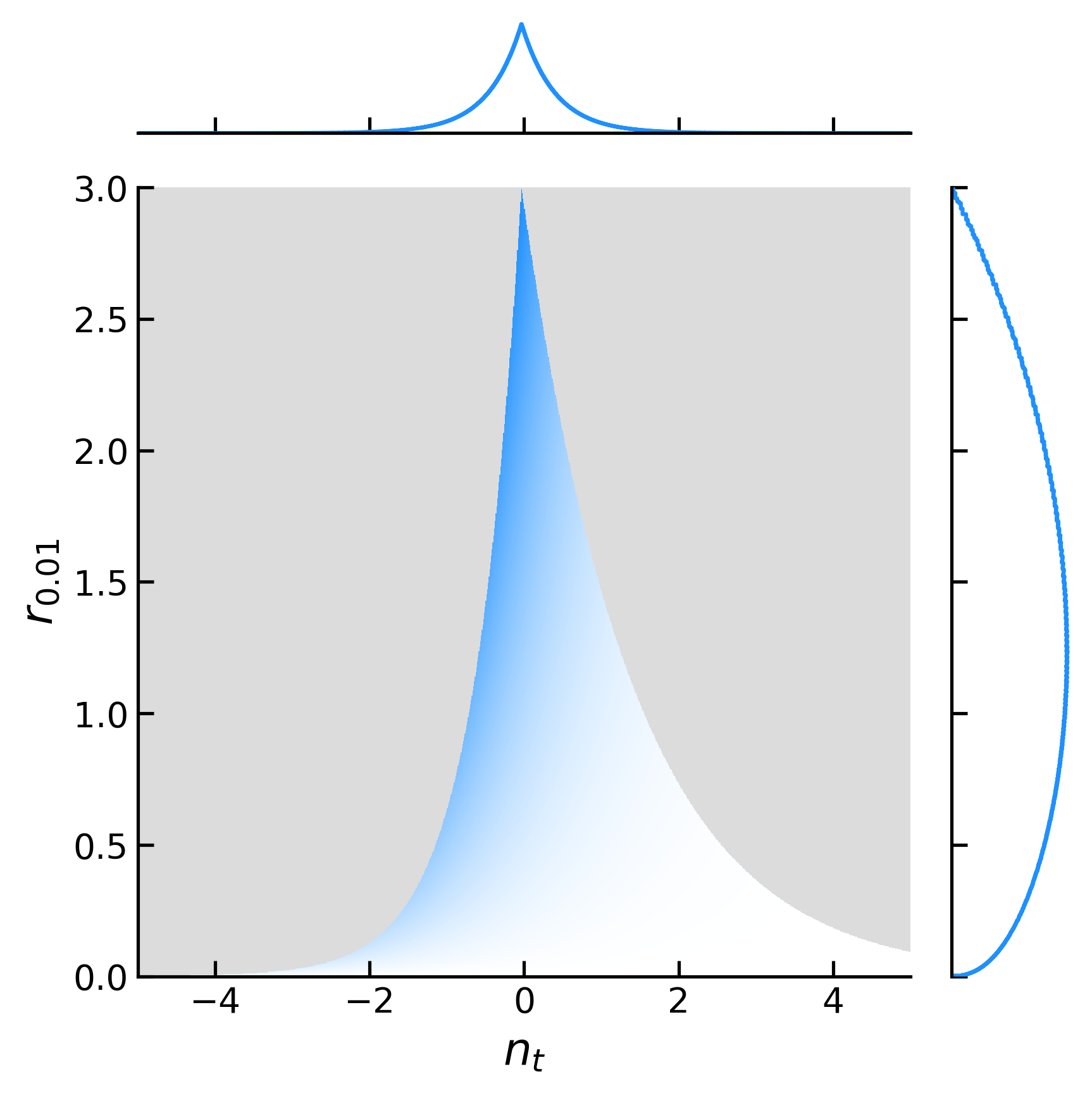}
    \hfill
    \includegraphics[width = .45\hsize]{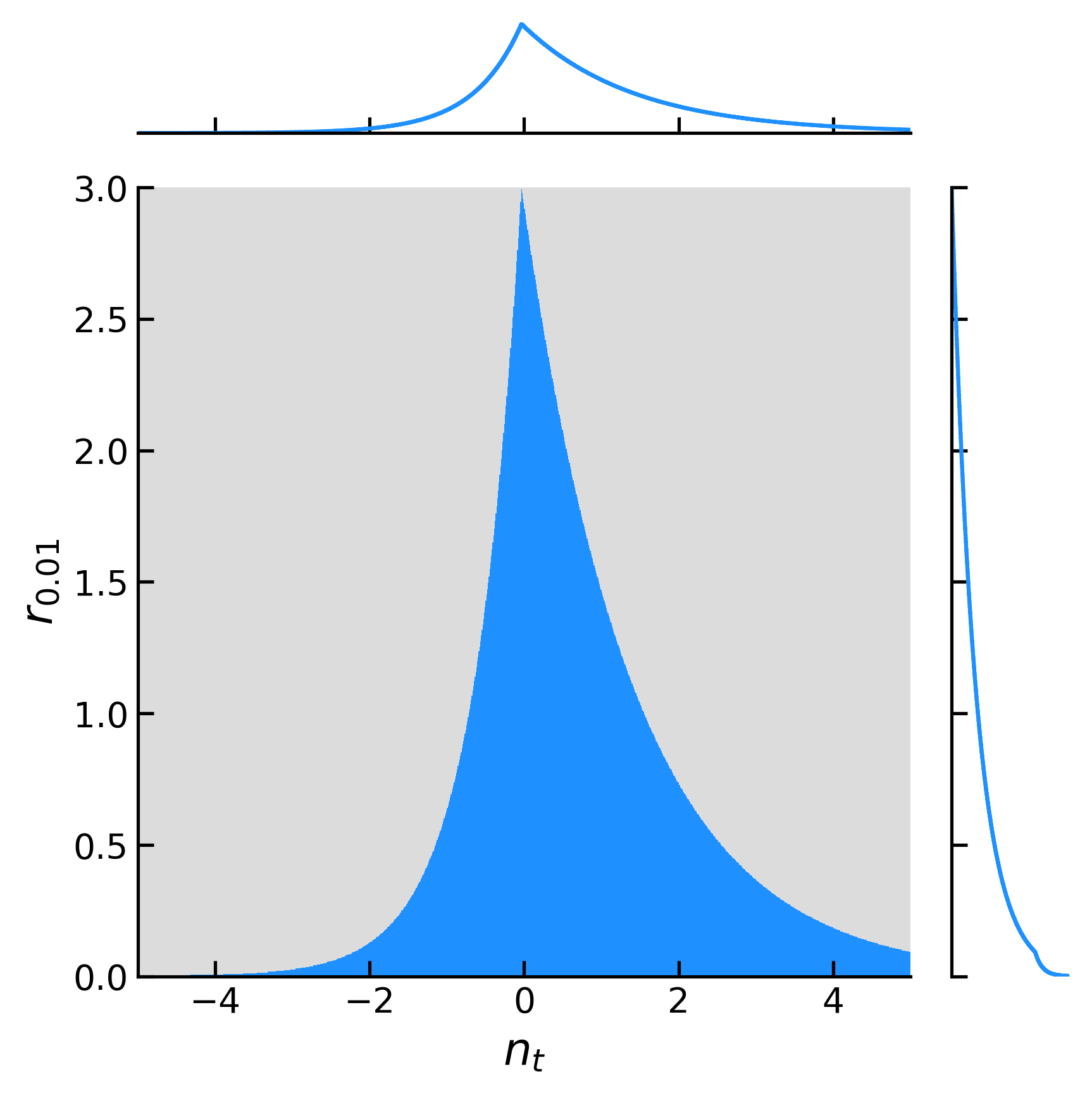}
    \caption{1D and 2D histograms on the $(r_{0.01}, n_t)$ plane, obtained from a uniform grid of points in the $\qty(r_1-r_2)$ plane. The left panel shows the simple result of the coordinate transformation, while the right one is the Jacobian reweighting.}
    \label{fig: analytical_priors}
\end{figure}

Now, given the nature of the MCMC analysis, one may ask himself if the results shown are flawed by the fact that the sampling is finite. Indeed, at some point, the MCMC converges and stops exploring the parameter space. To answer the question, we take here a more direct approach to the problem, which does not imply any sampling effect. Indeed, we can find the probability according to the TSA of a point in the $r_{0.01}-n_t$ plane given the prior probability distribution of the TSA on the $r_1-r_2$ plane (tab.\ref{tab: priors}) and recalling eq.\ref{eq: coordinate}. It reads
\begin{equation}
    \mathcal{P}^{\rm TSA}(r_{0.01}, n_t) = \eval{\mathcal{P}(r_1, r_2)}_{r_{1/2}=r_{1/2}(r_{0.01}, n_t)} \abs{\frac{d(r_1, r_2)}{d(r_{0.01}, n_t)}}\ .
\end{equation}
Here, $\mathcal{P}(r_1, r_2)$ acts as a ``selection function'': for each $r_{0.01}$ and $n_t$, it gives a constant factor if the corresponding $r_1$ and $r_2$ are within their prior ranges, zero otherwise. The second term is the inverse Jacobian of the coordinate transformation. As mentioned in sec. \ref{sec: test_prior}, because of its nature an MCMC cannot explore regions of very low probability, since it is a finite sampling. So, plotting $\mathcal{P}^{\rm TSA}$ we can have a look at the distribution to which the MCMC tends while the steps go to infinity. The left panel of figure \ref{fig: analytical_priors} shows this, while we color in gray the points where the probability is identically null. As you can see, the region at low $r_{0.01}$ and high $n_t$ has a negligible probability, which however is not zero. Instead, the gray region corresponds to the $r_{0.01}-n_t$ points bringing to $r_1-r_2$ points outside their prior.

We can now reweigh the obtained distribution. The weights are given by the ratio 
\begin{equation}
    \frac{\mathcal{P}^{\rm SSA}(r_{0.01}, n_t)}{\mathcal{P}^{\rm TSA}(r_{0.01}, n_t)} = \frac{\mathcal{U}(r_{0.01})\mathcal{U}(n_t)}{\mathcal{P}^{\rm TSA}(r_{0.01}, n_t)}\ ,
\end{equation}
where $\mathcal{U}(r_{0.01}),\mathcal{U}(n_t)$ are the uniform distributions reported in tab.\ref{tab: priors}. Thus, it is clear that the re-weighted probability goes back to a constant. Despite this, note that reweighting cannot recover completely the SSA prior, since a large portion of the SSA prior is excluded because it brings you outside of the $r_1-r_2$ priors (the gray region). For this reason, the bias carried by using the TSA can only be alleviated, but not solved, even when considering the theoretical probability distribution.

Compared with sec.\ref{sec: test_prior}, even if the 2D histograms seem quite different from the distributions shown in figure \ref{fig: priors} and \ref{fig: jacobian_scales}, we remind the reader that the contour plots are obtained by using a Gaussian kernel. If we apply a kernel to the histograms shown in figure \ref{fig: analytical_priors} we obtain very similar contours. 

On the other hand, note that the marginalized 1D histograms are very similar to those shown in \ref{fig: priors} and \ref{fig: jacobian_scales}. The $95\%$ CL intervals for the left panel are $0.15<r_{0.01}<2.68$ and $-1.33<n_{t}<1.27$ which are consistent with those of section \ref{sec: MCMC}. For what regards the right panel, we get $r_{0.01}<2.11$ and $-1.61<n_{t}<3.90$ at $95\%$ CL. Thus, as in section \ref{sec: MCMC}, the $2\sigma$ detection of $r_{0.01}$ is no longer there, while the pull toward scale invariance is just alleviated. Note that the upper bound on $n_t$ is the most affected by the finite sampling, as expected. Thus, if one wants to use the TSA to study some datasets and use reweighting to alleviate the bias, some extra attention must be given to the length of the sampling. The larger the better to represent the low-probability regions. Still randomly exploring these regions may be unfeasible. In general, the same could be said for any MCMC analysis, however, we find the TSA to be particularly sensitive to this.

\begin{figure}[tbp]
    \centering
    \includegraphics[width = .45\hsize]{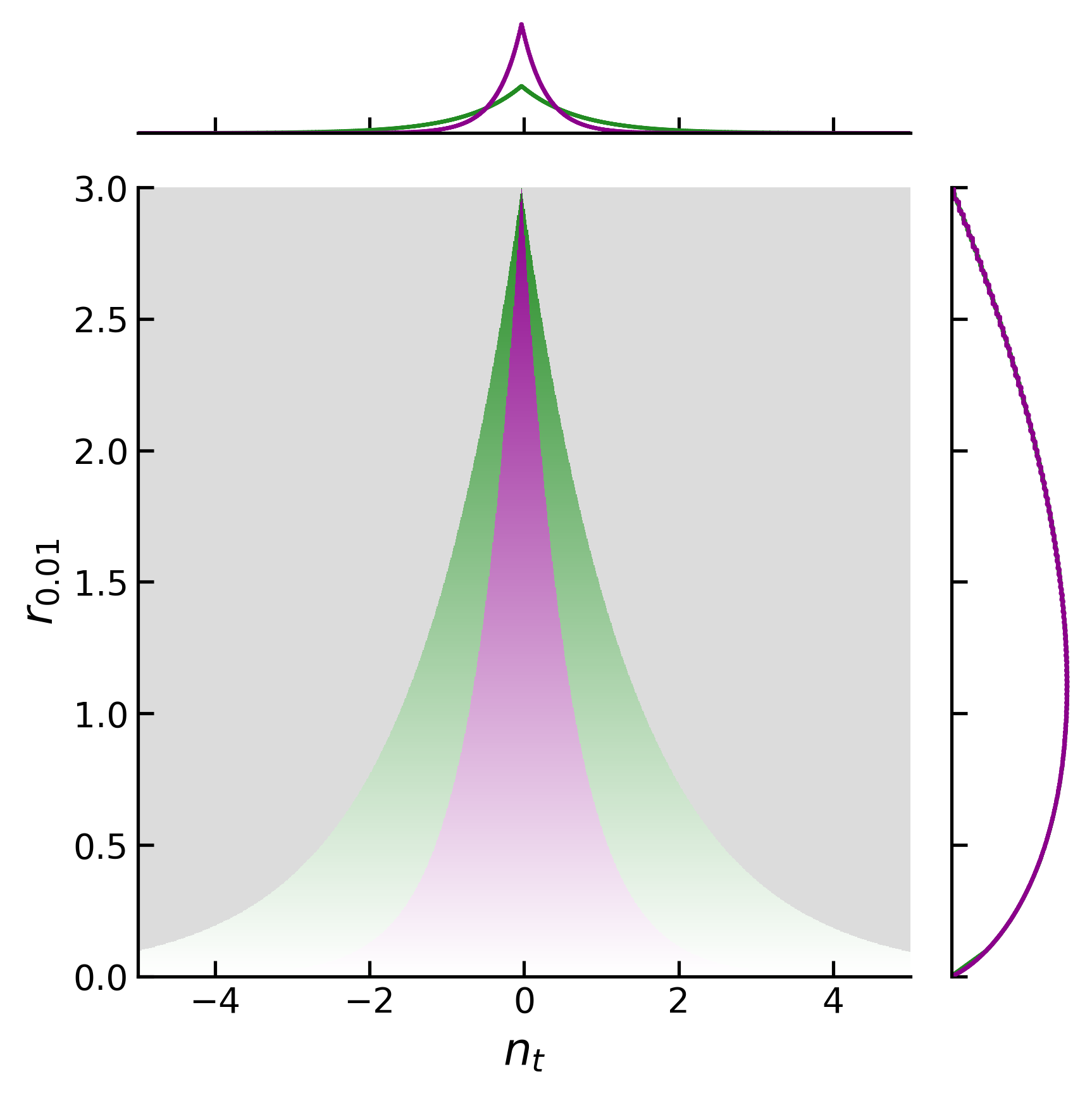}
    \caption{1D and 2D histograms on the $(r_{0.01}, n_t)$ plane, obtained from a uniform grid of point in the $\qty(r_1-r_2)$ plane when changing the leverage arm. The green one is obtained with $(k_1, k_2) = (0.005, 0.02)$ Mpc$^{-1}$ and the purple on with $(k_1, k_2) = (0.002, 0.05)$ Mpc$^{-1}$, mimicking section \ref{sec: MCMC}.}
    \label{fig: analytical_scales}
\end{figure}

With the same method, we can also show what happens when we change the lever arm $k_2-k_1$. As expected, the results shown in figure \ref{fig: analytical_scales} are similar to those in section \ref{sec: MCMC}.

In conclusion, this alternative way of exploring the prior information brought by the TSA leaves our conclusions of section \ref{sec: test_prior} unchanged.

\section{Results using the TSA} \label{sec: TSA}
As mentioned at the end of section \ref{sec: test_exact}, despite the choice we made to report our main results, we also performed the analysis using TSA. This allows us to appreciate the differences in the final posteriors due to the aspects underlined in section \ref{sec: MCMC}. Sticking to \citep{Planck_2018} we choose $(k_1,k_2) = (0.002, 0.02)$ Mpc$^{-1}$ and the priors shown in Tab.\ref{tab: priors}.

\begin{figure}[t]
    \centering
    \includegraphics[width=.49\hsize]{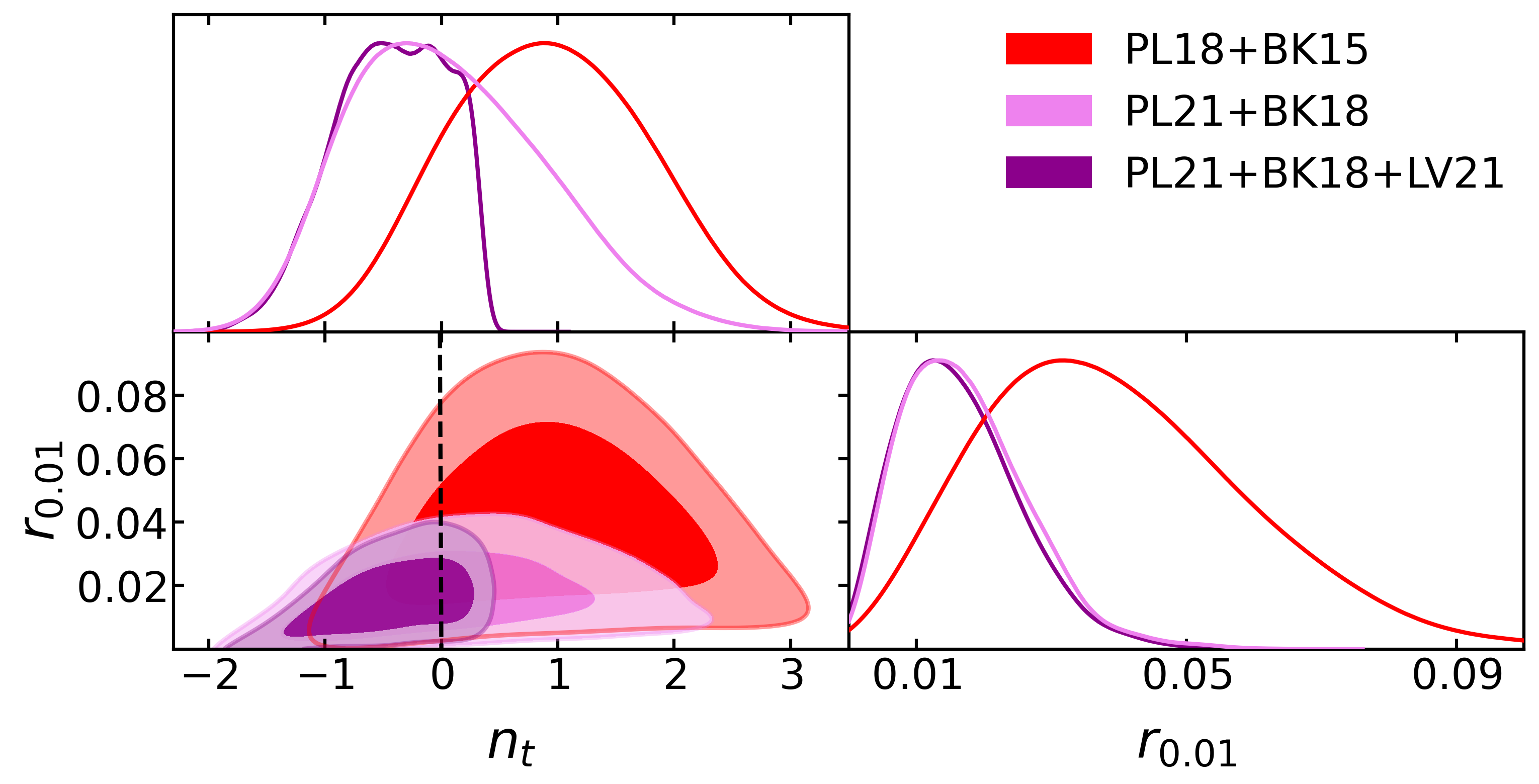}
    \hfill
    \includegraphics[width=.49\hsize]{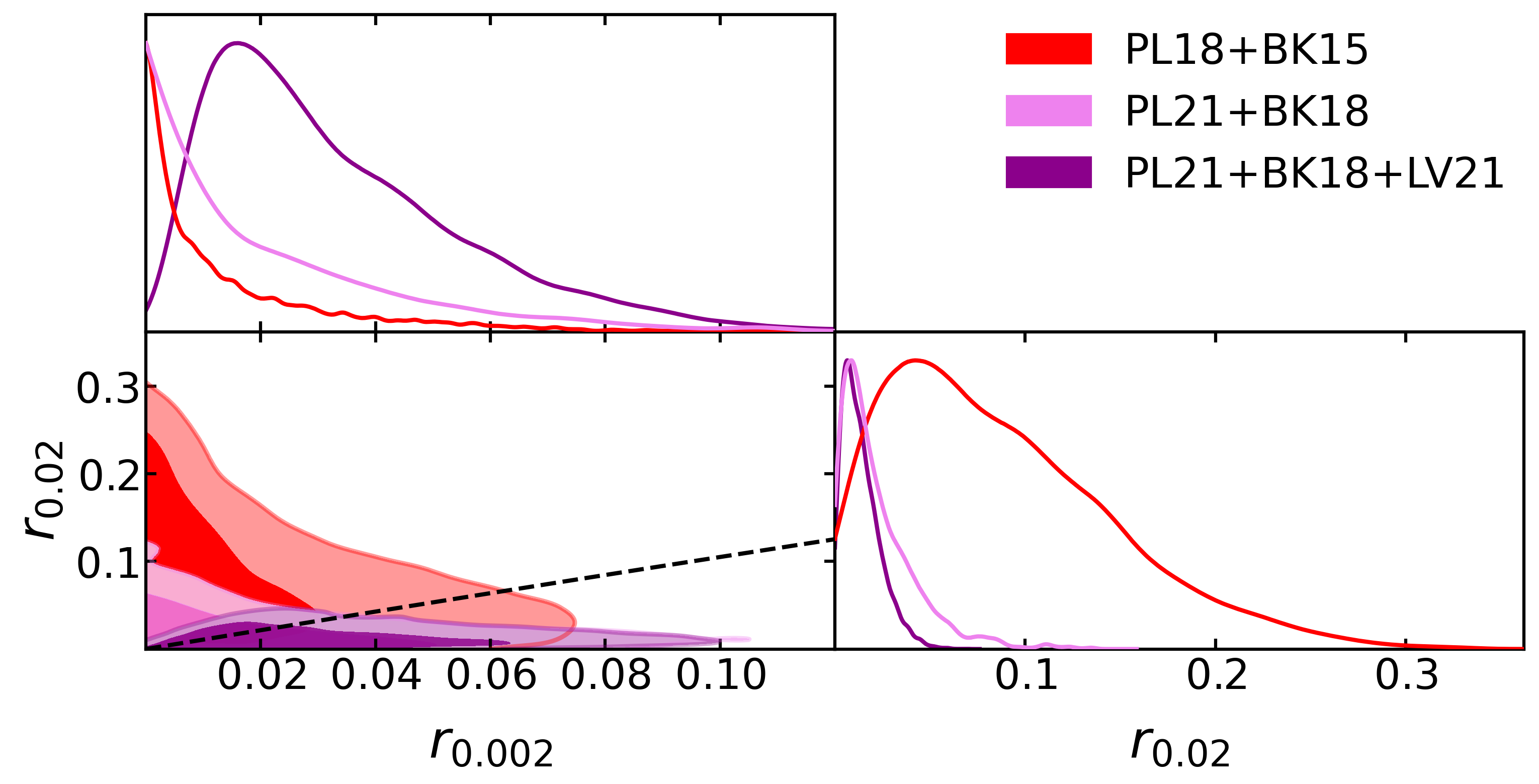}
    \caption{2D $68$ and $95\%$ CL intervals in the $\qty(r_{0.01}, n_t)$ plane (left) and the $\qty(r_1, r_2)$ plane (right) for PL18$+$BK15, PL21$+$BK18 and PL21$+$BK18$+$LV21 using the TSA. The dashed black line is the well-known slow-roll single-field prediction $n_t = -r/8 = -2 \epsilon$.}
    \label{fig: TSA_2D_res}
\end{figure}

Tab.\ref{tab: TSA_2D_res} reports the one-dimensional $95\%$ CL for each combination analyzed, together with the best fit, mean values, and the convergence parameters (see Tab.\ref{tab: 2D_res} for the SSA results).
\begin{table*}
\footnotesize
\centering
\addtolength{\leftskip} {-2cm}
\addtolength{\rightskip}{-2cm}
  \begin{tabular}{l , , .}
   \toprule
    & \multicolumn{1}{c}{$\mathbf{r_{0.01}}$ \textbf{95\% CL}}  & \multicolumn{1}{c}{$\mathbf{n_t}$ \textbf{95\% CL}} & \multicolumn{1}{c}{$\mathbf{R-1}$ \textbf{test}} \\ \midrule
   PL18+BK15       & [0.004,\ 0.079] & [-0.62,\ 2.63] & 0.008 \\
   PL18+BK18       & [0.003,\ 0.039] & [-1.07,\ 2.10] & 0.016 \\
   PL18+BK15+LV18  & [0.001,\ 0.070] & [-0.77,\ 0.42] & 0.010 \\
   PL18+BK15+LV21  & [0.001,\ 0.070] & [-0.79,\ 0.39] & 0.013 \\
   PL18+BK18+LV21  & [0.001,\ 0.035] & [-1.09,\ 0.39] & 0.019 \\
   PL21+BK15       & [0.002,\ 0.064] & [-1.06,\ 2.68] & 0.027 \\
   PL21+BK18       & [0.001,\ 0.035] & [-1.43,\ 1.74] & 0.035 \\
   PL21+BK18+LV21  & [0.001,\ 0.033] & [-1.30,\ 0.37] & 0.005 \\ \bottomrule
  \end{tabular}
\caption{95\% CL intervals of the 8 considered combinations of datasets. These have been obtained through the TSA. Here we also show the results of the Gelman-Rubin test for each combination.}
\label{tab: TSA_2D_res}
\end{table*}
In particular, we can notice that the most constraining is again PL21$+$BK18$+$LV21 (see section \ref{sec: data} for more details on the likelihoods we include in the analysis), which results in $0.001 < r_{0.01} < 0.033$ and $-1.30 < n_t < 0.37$ at $95\%$ CL. For this case, we show the 2D contours in figure \ref{fig: TSA_2D_res}, compared to some other relevant cases: PL18$+$ BK15 (representing the state of the art) and PL21$+$BK18 (to underline the improvement carried out by small-scale measurement).

Both Tab.\ref{tab: TSA_2D_res} and figure \ref{fig: TSA_2D_res} show that the TSA mimics a $2\sigma$-detection on every dataset we considered, while generally shrinking the bounds obtained on the tilt (see section \ref{sec: MCMC}). 

\begin{figure}[t]
    \centering
    \includegraphics[width=\textwidth]{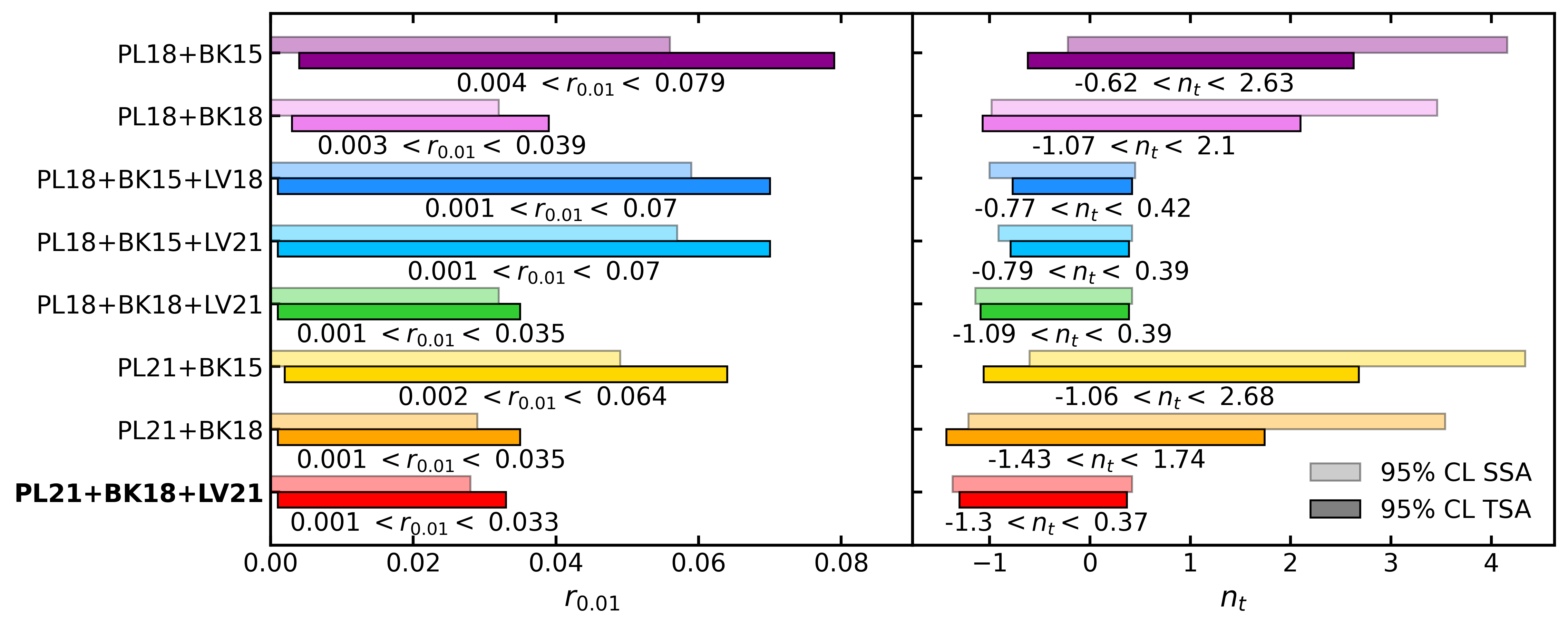}
    \caption{$95\%$ CL intervals for $r_{0.01}$ and $n_t$, considering different datasets, given in table \ref{tab: names}. Here we compare SSA and TSA. Our main result is PL21$+$BK18$+$LV21.}
    \label{fig: TSA_bounds}
\end{figure}

With figure \ref{fig: TSA_bounds}, one can compare the results on the $95\%$ CL intervals for $r_{0.01}$ and $n_t$ from SSA and TSA visually, mimicking figure \ref{fig: bounds}.

On a more technical note, the 95\% CL intervals shown in Tab.\ref{tab: TSA_2D_res} are obtained using \texttt{GetDist}, which will try to find the values of $r_{0.01}$ or $n_t$ between which the integral of the posterior gives 95\% of the total. It will attempt to do so in a two-tail fashion; if it fails, meaning that the upper or lower bound does not exist, it will shift to a single-tail. Thus, in the case of TSA, \texttt{GetDist} is capable of recovering a two-tailed bound consistently with every dataset. To avoid this, we would have to worsen the 1D and 2D smoothing scale parameters of \texttt{GetDist}, together with the approximation used by the package to account for the prior bounds (the keywords are ``boundary\_correction\_order'' and ``mult\_bias\_correction\_order'', which are left by default to first-order correction \footnote{For more details on these parameters, see \url{https://getdist.readthedocs.io/en/latest/index.html} .}).

Furthermore, we can compare the first row of table \ref{tab: 2D_res} with the results of \textit{Planck} 2018 \cite{Planck_2018}, mentioned in section \ref{sec: state}. \textit{Planck} reports $r_{0.01}<0.076$ and $-0.55 < n_t < 2.54$ at $95\%$ CL, whereas we obtain a $2\sigma$ detection on $r_{0.01}$ and slightly wider bounds on $n_t$. As mentioned above, if we set the 1D and 2D smoothing scale parameters to 0.35 and 0.40 (these are units of the width of the \texttt{GetDist} bins), respectively, and the approximations to account for the prior bounds to their minimum, we recover the exactly the bound on the amplitude $r_{0.01}<0.076$, while we obtain a larger bound on the tilt, that is $-0.79 < n_t < 2.68$. This residual difference in the tilt may be due to a different level of convergence between the two analyses. Note that we run our MCMC very deep into convergence to ensure a careful exploration of the tails of the distributions.

\begin{figure}[t]
    \centering
    \includegraphics[width = .49\textwidth]{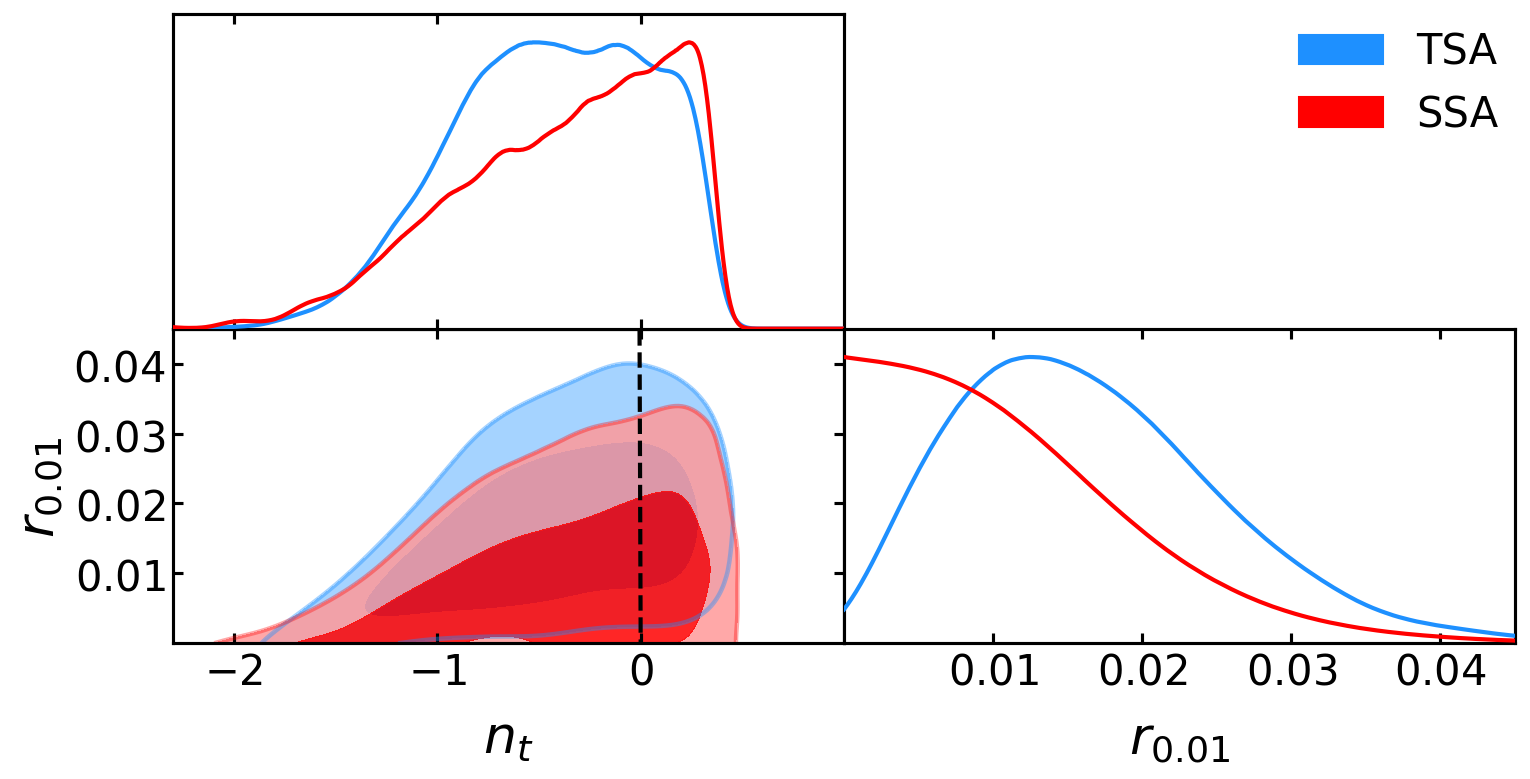}
    \caption{1D and 2D posteriors on the $r_{0.01}$ and $n_t$, obtained using the SSA and the TSA on the PL21+BK18+LV21 dataset. The dashed lines represent the single-field slow-roll prediction for $n_t = -2\epsilon$.}
    \label{fig: comparison_data}
\end{figure}

In figure \ref{fig: comparison_data} we also show the comparison between the posterior distributions obtained using the SSA and the TSA on the PL21+BK18+LV21 dataset.

\end{appendix}

\acknowledgments

The authors thank Adrià Gómez-Valent, Eiichiro Komatsu, Fabio Finelli, Luca Pagano, Mario Ballardini, Martina Gerbino, Massimiliano Lattanzi, and Matthieu Tristram for useful comments and discussions. This work is based on observations obtained with \textit{Planck} (http://www.esa.int/Planck), an ESA science mission with instruments and contributions directly funded by ESA Member States, NASA, and Canada. Some of the results in this paper have been derived using the following packages: \texttt{CAMB} \cite{Lewis:2013hha, Lewis_2000, Howlett_2012}, \texttt{Cobaya} \cite{Torrado:2020xyz}, \texttt{GetDist} \cite{Lewis:2019}, \texttt{Matplotlib}\footnote{\url{https://github.com/matplotlib/matplotlib}.} \cite{matplotlib} and \texttt{NumPy}\footnote{\url{https://github.com/numpy/numpy}.} \cite{numpy}. 
G.G., M.M., N.B., N.V., and S.M. acknowledge support from the COSMOS network (\url{www.cosmosnet.it}) through the ASI (Italian Space Agency) Grants 2016-24-H.0, 2016-24-H.1-2018 and 2020-9-HH.0. A.R.~acknowledges funding from MIUR through the ``Dipartimenti di eccellenza'' project Science of the Universe. We acknowledge partial support from INFN through the InDark initiative.

% \paragraph{Note added.} This is also a good position for notes added
% after the paper has been written.


% \input{biblio.bbl}

% \bibliographystyle{plainnat}
% \bibliographystyle{style_bib}
% \bibliographystyle{unsrtnat}
% \bibliographystyle{JHEP}

\bibliographystyle{apsrev}
\bibliography{bibliography}

\end{document}